\documentclass[twocolumn]{aastex631}

\usepackage{amsmath}
\usepackage{soul}

\newcommand{\cH}{\mathcal{H}}

\shorttitle{Water-world atmospheres}
\shortauthors{Kempton et al.}

\graphicspath{{./}{figures/}}

\begin{document}

\title{Where are the Water Worlds?: Self-Consistent Models of Water-Rich Exoplanet Atmospheres}

\author[0000-0002-1337-9051]{Eliza M.-R. Kempton}
\affiliation{Department of Astronomy, University of Maryland, College Park, MD 20742, USA}

\author[0000-0001-9939-5564]{Madeline Lessard}
\affiliation{Department of Astronomy, University of Maryland, College Park, MD 20742, USA}

\author[0000-0002-2110-6694]{Matej Malik}
\affiliation{Department of Astronomy, University of Maryland, College Park, MD 20742, USA}

\author[0000-0003-0638-3455]{Leslie A. Rogers}
\affiliation{Department of Astronomy \& Astrophysics, University of Chicago, 5640 S Ellis Ave, Chicago, IL 60637, USA}

\author{Kate E. Futrowsky}
\affiliation{Department of Astronomy, University of Maryland, College Park, MD 20742, USA}

\author[0000-0003-2775-653X]{Jegug Ih}
\affiliation{Department of Astronomy, University of Maryland, College Park, MD 20742, USA}

\author{Nadejda Marounina}
\affiliation{Department of Astronomy \& Astrophysics, University of Chicago, 5640 S Ellis Ave, Chicago, IL 60637, USA}

\author[0000-0001-7152-9794]{Carlos E. Mu\~{n}oz-Romero}
\affiliation{Center for Astrophysics $\mid$ Harvard \& Smithsonian, 60 Garden Street, Cambridge, MA 02138, USA}
\affiliation{Department of Physics, Grinnell College, 1116 8th Ave., Grinnell, IA 50112, USA}

\begin{abstract}

It remains to be ascertained whether sub-Neptune exoplanets primarily possess hydrogen-rich atmospheres or whether a population of H$_2$O-rich “water worlds” lurks in their midst. Addressing this question requires improved modeling of water-rich exoplanetary atmospheres, both to predict and interpret spectroscopic observations and to serve as upper boundary conditions on interior structure calculations. Here we present new models of hydrogen-helium-water atmospheres with water abundances ranging from solar to 100\% water vapor. We improve upon previous models of high water content atmospheres by incorporating updated prescriptions for water self-broadening and a non-ideal gas equation of state. Our model grid (\url{https://umd.box.com/v/water-worlds}) includes temperature-pressure profiles in radiative-convective equilibrium, along with their associated transmission and thermal emission spectra.  We find that our model updates primarily act at high pressures, significantly impacting bottom-of-atmosphere temperatures, with implications for the accuracy of interior structure calculations.  Upper atmosphere conditions and spectroscopic observables are less impacted by our model updates, and we find that under most conditions, retrieval codes built for hot Jupiters should also perform well on water-rich planets.  We additionally quantify the observational degeneracies among both thermal emission and transmission spectra.  We recover standard degeneracies with clouds and mean molecular weight for transmission spectra, and we find thermal emission spectra to be more readily distinguishable from one another in the water-poor (i.e.\ near-solar) regime.

\end{abstract}



\section{Introduction} \label{sec:intro}

The nature of the population of exoplanets with radii between 1.5 and 4 $R_\oplus$ remains enigmatic.  Statistically speaking, such planets are unlikely to be rocky in nature \citep{weiss14,rogers15,fulton17}, and by their sizes and masses alone, these planets also don't fit cleanly into the jovian vs.\ terrestrial classification scheme of our solar system.  Instead, these so-called ``sub-Neptunes" reside in the nexus of degenerate mass-radius parameter space with regard to inferring their bulk compositions.  The bulk densities of such planets tend to be equally well matched by a solid rock/iron core and a gaseous hydrogen-rich atmosphere, or an icy water-rich interior and atmosphere \citep[e.g.][]{adams08, rogers10a, rogers10, valencia13}.  The former scenario we refer to in this paper as a ``gas dwarf", whereas the latter we term as a ``water world". 

From a planet formation standpoint, both gas dwarfs and water worlds are potential outcomes of low-mass planet assembly.  In the former case, the planets primarily form close-in to their host stars from ice-poor pebbles, prior to accreting a primordial atmosphere from the gaseous disk \citep[e.g.][]{Hansen&Murray2012ApJ, Chiang&Laughlin2013MNRAS, Bodenheimer&Lissauer2014ApJ, lee16}.  If the accreted atmosphere is too massive, additional loss mechanisms are required to reduce the planet to its present-day size \citep[e.g.][]{owen17,ginzburg18,wyatt20}.  A water world outcome instead implies a very different birth scenario, likely involving formation beyond the snow line or significant pollution by planetesimals from that region \citep{Kuchner2003ApJL, LegerEt2004Icarus, bitsch19, izidoro21}.

To determine which scenario best describes the population of observed sub-Neptunes, astronomers of late have invoked population-level statistical analyses.  Specifically, theories have been developed that attempt to recover the bi-modal radius distribution of planets smaller than $\sim4$ $R_\oplus$, as well as dependencies of that distribution on additional variables such as orbital distance and stellar host type.  Among these theories, both photoevaporation \citep{owen17, rogers21} and core-powered mass loss \citep{ginzburg18} explain the planet radius distribution of Kepler-discovered planets orbiting F/G/K stars with mass loss from an accreted H$_2$ atmosphere overlying a rocky planetary core --- giving support to the gas-dwarf hypothesis.  When low-mass host stars are investigated separately, a different evolution of the radius gap with irradiation is uncovered \citep{cloutier20}, potentially indicating an additional formation/evolution mechanism for sub-Neptunes orbiting later-type stars.  This second mechanism, perhaps gas-poor formation \citep{lee14,lee16}, also points to sub-Neptunes being gas dwarfs, albeit via a different formation pathway.  Based on the success of these three theories (photoevaporation, core-powered mass loss, and perhaps gas-poor formation)\footnote{A fourth theory --- impact erosion \citep{wyatt20} --- has also been invoked to explain the existence of the radius gap, in further support of the gas dwarf hypothesis.  This theory has yet to be applied though to formally reproduce the observed sub-Neptune / super-Earth size distribution.} to explain the radius distribution of known exoplanets, the prevailing view currently is that sub-Neptunes are primarily gas dwarfs  \citep[e.g.\ see a recent review by][]{bean21}.

The existence of water worlds within the sub-Neptune population is certainly not ruled out though.  Multiple individual planets possess bulk densities that are fully consistent with a water-rich composition \citep[e.g.][]{valencia13,acuna22,piaulet23}.  Furthermore, on the population level, \citet{zeng19}, \citet{venturini20}, \citet{NeilEt2022ApJ}, and \citet{luque22} all find evidence that ice-rich water worlds are embedded in the sub-Neptune population by comparing interior structure and planet evolution models to Kepler radius statistics and mass-radius measurements.
\citet{dorn21} and \citet{mousis20} furthermore show that interior structure models that include additional physics (i.e.\ a ``wet melt" transition or a supercritial hydrosphere) can imply much higher water abundances for a planet with a given mass and radius.

A growing body of literature on magma-atmosphere interactions has the potential to further complicate matters.  For example, \citet{kite21} predict that planets just above the radius gap may have thick ($10-2000$ bar) H$_2$O atmospheres overlying bulk rocky (rather than icy) interiors,  creating the possibility of a third class of of sub-Neptune that is neither gas dwarf nor water world.   Conversely, \citet{madhu21} expand on the existence of water-rich planets to implications of habitability, showing that moderately irradiated sub-Neptunes with hydrogen-rich envelopes may harbor sub-surface liquid oceans and water-rich interiors underneath their thick atmospheres.  Regardless, it is clear that water-rich planets and planetary envelopes remain a viable possibility among the known population of sub-Neptunes.

To resolve the gas dwarf vs.\ water world ambiguity, atmospheric observations have been proposed as an avenue for breaking degeneracies in sub-Neptune bulk composition \citep{millerricci09,millerricci10,rogers10}.  However, observations of sub-Neptune atmospheres to-date have generally brought inconclusive evidence due to muted or absent spectral features, perhaps indicating pervasive aerosols \citep[e.g.][]{bean10,kreidberg14,crossfield17,guo20}.  This is expected to change in the upcoming era of precision spectroscopy with the James Webb Space Telescope (JWST).  The JWST observatory is already scheduled to characterize many sub-Neptune exoplanets in its first cycle of community and guaranteed-time observations\footnote{A full list of transiting exoplanet observations with JWST is provided here: \url{https://tess.mit.edu/science/tess-acwg/##transiting-exoplanet-observations-jwst}}.  Such observations are anticipated to have the precision and wavelength coverage to break aerosol-composition degeneracies, allowing astronomers to constrain atmospheric water abundances directly.  

Previous models of sub-Neptune exoplanet atmospheres have focused on a range of hydrogen-rich, metallicity-enhanced, and high mean molecular weight scenarios \citep{millerricci09, millerricci10, howe12, benneke12, benneke13, morley13, morley15, hu14, madhu15, kawashima18, piette20}.  Those studies that have examined water-rich atmospheres have typically done so using models intended for gas giant planets as the basis of their calculations --- specifically, the water opacities employed were calculated assuming a background atmosphere of H$_2$ or occassionally Earth-like air, rather than water vapor.  

Furthermore, most studies of observable traits have focused on transmission spectroscopy because it is perceived that this will be the dominant observational technique used to characterize sub-Neptunes.  However, thermal emission spectroscopy (i.e.\ obtained during secondary eclipse) has increased sensitivity to vertical thermal structure, decreased sensitivity to clouds, and produces higher signal-to-noise for the hottest planets.  Far less attention has been paid to modeling sub-Neptune emission spectra in the literature to date. Of note, no published works thus far have calculated thermal profiles of water-dominated atmospheres in radiative-convective equilibrium --- parameterized and/or otherwise simplified (e.g.\ double-gray) models have instead been the norm \citep[e.g.][]{millerricci10,benneke13,madhu15}.

In this paper we focus on missing aspects of the exoplanet atmosphere modeling toolkit that may impact the ability to robustly infer atmospheric water content from exoplanetary spectra, especially in the case of very water-rich atmospheres.  We generate a large model grid of water-hydrogen-helium atmospheres in radiative-convective equilibrium with bulk water abundances ranging from solar up to 100\% water vapor.  We include updated treatments for opacities and convective lapse rates in the water-rich regime, and quantify the impacts of these model improvements.  We also search for unexplored degeneracies in interpreting emission spectra of water-rich atmospheres.  We limit ourselves in the current work to combinations of water, hydrogen, and helium only, in order to not complicate our analysis with additional spectroscopically active species.  The framework that we develop here can readily accommodate the addition of trace and non-trace components of other molecules though --- an extension that should be pursued in future work.  Similarly, we also limit our study to planets interior to the habitable zone to avoid modeling complications related to water condensation, but in future work we intend to address habitable zone exoplanets with water clouds and surface oceans.  

Our water-hydrogen-helium atmosphere models serve as a basis for interpreting sub-Neptune spectra in the era of JWST, as upper boundary conditions for planetary interior models, and as a framework for self-consistently incorporating high mean molecular weight effects into exoplanet atmosphere models that otherwise more typically simulate hydrogen-dominated conditions.  
In Section~\ref{sec:methods} of this paper we describe our atmosphere model grid and techniques, in Section~\ref{sec:results} we present our results, and we discuss and conclude in Section~\ref{sec:conclusion}.

\section{Methods} \label{sec:methods}

\subsection{Model Grid} \label{sec:grid}

We construct our baseline grid of model atmospheres as a function of water (volumetric) mixing ratio, internal temperature, $T_{int}$, and surface gravity, $g$.  The model parameters considered are given in Table~\ref{tab:main_grid}. The selected parameter range is motivated by several considerations: (1) we aim to span the parameter space of observationally accessible sub-Neptunes, (2) we wish to consider atmospheric compositions that are consistent with both accreted solar composition gas and ice-dominated worlds, and (3) for interfacing with interior structure models, we must cover the full range of low surface gravity and high $T_{int}$ values that are experienced by young sub-Neptunes that ultimately evolve into the planets that we observe today.  In total, this baseline model grid spans 16 H$_2$O mixing ratios, 14 values of $T_{int}$, and 15 values of $\log g$ for a total of 3,360 individual atmospheric models.  We furthermore consider two different levels of planetary irradiation --- specifically, equilibrium temperatures ($T_{eq}$) of 500~K and 700~K, assuming zero albedo and planet-wide heat redistribution --- and we model both an M-dwarf and a Sun-like host star.  This expands our baseline grid by an additional factor of 4 for a total of 13,440 model atmospheres.  
\begin{deluxetable}{ccccc}
\tablenum{1}
\tablewidth{0pt}
\tablecaption{Main model parameter grid\label{tab:main_grid}}
\tablehead{
\colhead{H$_2$O vol. mix. ratio} & \colhead{$T_{int}$} & \colhead{log$(g)$} & \colhead{$T_{eq}$} & \colhead{star type} \\
\colhead{} & \colhead{(K)} & \colhead{(cgs)} & \colhead{(K)} & \colhead{}
}
\startdata
0.00100 & 10 & 2.0 & 500 & M-dwarf \\
0.00158 & 20 & 2.3 & 700 & solar \\
0.00251 & 30 & 2.5 &    &    \\
0.00398 & 40 & 2.6 &    &    \\
0.00631 & 50 & 2.7 &    &    \\
0.01000 & 60 & 2.8 &    &    \\
0.01585 & 70 & 2.9 &    &    \\
0.02512 & 80 & 3.0 &    &    \\
0.03981 & 90 & 3.1 &    &    \\
0.06310 & 100 & 3.2 &    &    \\
0.10000 & 110 & 3.3 &    &    \\
0.15849 & 120 & 3.4 &    &    \\
0.25119 & 200 & 3.5 &    &    \\
0.39811 & 400 & 3.6 &    &    \\
0.63096 &    & 3.7 &    &    \\
1.00000 &    &    &    &    \\
\enddata
\tablecomments{Main model parameter grid. Emission spectra are modeled using both star types, while transmission spectra are only modeled using the M-dwarf star. All models in this main grid are run for cloud-free atmospheres and a GJ 1214 b-analog planet radius.}
\end{deluxetable}

For each model atmosphere, we generate both a one-dimensional (1-D) temperature-pressure (T-P) profile in radiative-convective equilibrium and a planetary emission spectrum consisting of emitted and reflected light components, as described below in Section~\ref{sec:T-P}.  We furthermore generate transmission spectra for each of the M-dwarf models as detailed in Section~\ref{sec:transmission}.  (We do not calculate transmission spectra for sub-Neptunes orbiting the Sun-like host star because the signal-to-noise for atmospheric characterization of such systems is expected to be out of observational reach.)  

We additionally calculate several sub-grids of models aimed at probing some of the assumptions made in our baseline modeling.  For instance, in our transmission spectrum modeling, we explore a range of cloudy atmospheres, detailed in Table~\ref{tab:cloudy_grid}.  All of our cloudy models are generated for a planet with $\log(g) = $ 2.9 \mbox{($g = 7.9$ m s$^{-2}$)} and $T_{int} = $ 40 K --- values that are selected to approximately align with the properties of the benchmark sub-Neptune GJ 1214b.  

We furthermore calculate a sub-grid of pure H$_2$ -- H$_2$O models for the Sun-like host star and 500~K equilibrium temperature case, which do not include any helium.  In contrast, our baseline grid employs a solar composition mix of hydrogen and helium for the non-H$_2$O component of each of the modeled atmospheres.  The aim of this sub-grid is to quantify the effect of an unconstrained helium mixing ratio on the outcomes of our atmosphere modeling\footnote{We do not show the models from this sub-grid later in the paper because the results are unremarkable.  The removal of He has no impact on the resulting T-P profiles or thermal emission spectra at an observable level.  The no-helium sub-grid models can be found in the online repository with the rest of our model grid, for completeness}.

Finally, we calculate a small sub-grid of models to assess the impact of our novel treatments of water opacities and convective lapse rates in our radiative transfer calculations.  For the water opacities, we specifically compare the treatment in our current work (described in Appendix~\ref{app:water_opac}) against the self-broadened H$_2$O opacities from \citet{gharib19}, against the self-broadened H$_2$O opacities from HITEMP \citep{rothman10}, and against the \mbox{H$_2$-broadened} water opacities that we have employed in our previous work. Note that we have found that the pre-tabulated, self-broadened opacity for 100\% H$_2$O provided by \citet{gharib19} is erroneous for $P \geq 3$ bar and $T \geq 1300$ K (confirmed by E.~Gharib-Nezhad, priv.~comm.). Hence, while using their opacity in our modeling, for $P \geq 3$ bar we extrapolate the opacity calculated at T = 1200 K. For the lapse rates, we compare our current calculations (see Appendix~\ref{app:adiabatic}) against those using constant lapse rates appropriate for ideal gases --- both diatomic (e.g.\ pure H$_2$) and triatomic (e.g.\ pure H$_2$O) --- as are typically employed in many exoplanet forward models.  The model parameters for this sub-grid are given in Table~\ref{tab:treatment_grid}, which amount to an additional 1152 modeled emission and transmission spectra and 288 T-P profiles.  The details of these model comparisons are expounded upon in greater detail in Section~\ref{sec:treatment_results}.


\begin{deluxetable}{ccc}
\tablenum{2}
\tablecaption{Cloudy model parameter grid\label{tab:cloudy_grid}}
\tablewidth{0pt}
\tablehead{
\colhead{H$_2$O vol. mix. ratio} & \colhead{Cloud-top pressure} & \colhead{$T_{eq}$} \\
\colhead{} & \colhead{(bar)} & \colhead{(K)}
}
\startdata
0.00100 & $10^0$ & 500 \\
0.00158 & $10^{-1}$ & 700 \\
0.00251 & $10^{-2}$ &    \\
0.00398 & $10^{-3}$ &    \\
0.00631 & $10^{-4}$ &    \\
0.01000 & $10^{-5}$ &    \\
0.01585 & $10^{-6}$ &    \\
0.02512 & clear &    \\
0.03981 &    &    \\
0.06310 &    &    \\
0.10000 &    &    \\
0.15849 &    &    \\
0.25119 &    &    \\
0.39811 &    &    \\
0.63096 &    &    \\
1.00000 &    &    \\
\enddata
\tablecomments{These models are all transmission spectra run for a GJ 1214b analog planet with \mbox{$g = 7.9$ m s$^{-2}$}, $T_{int} = 40$ K, and \mbox{$R_p = 0.238$ $R_{Jup}$}, orbiting an M-dwarf host star.  The clouds are added \textit{post hoc} on top of a T-P profile that is calculated for clear atmosphere conditions.}
\end{deluxetable}

\begin{deluxetable*}{cccccc}
\tablenum{3}
\tablecaption{H$_2$O treatment assessment parameter grid\label{tab:treatment_grid}}
\tablewidth{0pt}
\tablehead{
\colhead{$T{int}$} & \colhead{log$(g)$} & \colhead{$T_{eq}$} & \colhead{iterative opacity\tablenotemark{a}} & \colhead{post-process opacity\tablenotemark{b}} & \colhead{convective treatment} \\
\colhead{(K)} & \colhead{(cgs)} & \colhead{(K)} & \colhead{} & \colhead{} & \colhead{}
}
\startdata
40 & 2.0 & 500 & traditional\tablenotemark{c} & traditional & diatomic ideal gas \\
100 & 2.9 & 700 & ExoCross & ExoCross & triatomic ideal gas \\
200 & 3.7 &    & HITEMP (J-dependent) & HITEMP (J-dependent) & this work   \\
400 &    &    &  this work  & this work  &    \\
\enddata
\tablecomments{Model sub-grid to assess the impact of our novel H$_2$O opacity and lapse rate treatments.  For the T-P profile calculations all combinations of parameters in columns 1-4 and 6 are inter-compared (288 models).  For the emission and transmission spectra, all combinations are compared (1152} models).
\tablenotetext{a}{Water opacity treatment employed in the \texttt{HELIOS} radiative-convective equilibrium calculation.}
\tablenotetext{b}{Water opacity treatment employed in the post-processed \texttt{HELIOS} emission spectrum calculation.}
\tablenotetext{c}{H$_2$-broadened opacities, as typically employed in most exoplanet retrieval codes and giant planet forward models.}
\end{deluxetable*}

\subsection{\texttt{HELIOS} Temperature-Pressure Profiles and Emission Spectra \label{sec:T-P}}

We use the open-source 1-D radiative transfer code \texttt{HELIOS} \citep{malik19a, malik19b} for the numerical modeling of the atmospheric structure. For each combination of chemical and physical input parameters (Table~\ref{tab:main_grid}) we simulate the 1-D atmospheric T-P profile in radiative-convective equilibrium, assuming that heat is perfectly distributed over both hemispheres (i.e., heat redistribution factor 0.25), and the corresponding planetary emission spectrum. 

For modeled planets orbiting the solar-type star, the stellar flux is approximated by blackbody emission using the temperature $T_{\star} = 5800$ K and a solar radius, i.e., $R_\star = R_\sun$. For the M-dwarf host we use a PHOENIX stellar spectrum \citep{husser13}, interpolated for $T_{\star} = 3026$ K, $R_\star = 0.216$  $R_\sun$, $\log g_\star = 4.944$ and $[M/{\rm H}]_\star = -0.39$, consistent with GJ 1214 \citep{harpsoe13}. In all cases, the specified planetary equilibrium temperature (i.e.\ 500 K or 700 K) is achieved by setting the orbital distance in \texttt{HELIOS} to provide the correct value of $T_{eq}$, assuming zero albedo and full-planet heat redistribution.  


The H$_2$O opacity is calculated with the open-source code \texttt{HELIOS-K} \citep{grimm15, grimm21} using the POKAZATEL line list \citep{polyansky18}. The calculation uses a high resolution of 0.01~cm$^{-1}$ (which corresponds to $R=10^6$ at 1 $\mu$m) and assumes a Voigt profile for the spectral line shape. In addition to the standard H$_2$/He broadening coefficients provided by the POKAZATEL line list, we apply a novel treatment to account for the proper strength of water self-broadening depending on the atmospheric water fraction (see Appendix~\ref{app:water_opac} for details). Furthermore, we include collision-induced absorption (CIA) of H$_2$-H$_2$ and H$_2$-He pairs \citep{borysow02, richard12} and Rayleigh scattering of H$_2$O \citep{cox00, wagner08}, H$_2$ \citep{cox00} and He \citep{sneep05, thalman14}. 

The radiative transfer calculation is performed as a two-step process. First, in order to find the equilibrium T-P profile, we use the k-distribution method with 20 Gaussian points in each of 410 wavelength bins between 0.244 $\mu$m and $10^5 \mu$m. The k-distribution method takes the high-resolution opacity into account, preserving the physically accurate transmission within each bin. Since in this work we only consider absorption lines from a single gas species, H$_2$O, the accuracy of the k-distribution method does not suffer from opacity mixing approximations (such as the correlated-k assumption). Second, once the model is converged, we generate the planet's emission spectrum (including both thermal emission and reflected light components) from the previously calculated T-P profile, using opacity sampling, re-binned from the initial opacity resolution down to $R$ = 3000. For each case, k-distribution method and sampling, we use premixed opacity tables that include all extinction coefficients weighted by their respective atmospheric abundance.

Convection is treated in \texttt{HELIOS} via a convective adjustment. This numerical scheme requires the adiabatic coefficient (determining the lapse rate in convective regions) and the specific heat capacity. In contrast to previous works \citep[e.g.,][]{malik19b}, for our baseline modeling we do not assume any ideal gas approximations. We instead calculate both quantities directly from the specific entropy of the gas, taking the local gas composition into account (see Appendix~\ref{app:adiabatic} for details).

All of the \texttt{HELIOS} T-P profiles are calculated assuming cloud-free conditions.  Since water is a major component of each of these atmospheres, we check each of the resulting T-P profiles against conditions for water to condense into the liquid or solid phase.  All of the atmospheres in our model grids remain in the gas-phase region of parameter space for all atmospheric constituents.

\subsection{\texttt{Exo-Transmit} Transmission Spectra \label{sec:transmission}}

We use the \texttt{Exo-Transmit} radiative transfer code \citep{kempton17} to generate transmission spectra for each of the simulated planets orbiting the M-dwarf host star.  To specify the thermal structure of the atmosphere we use the \texttt{HELIOS} T-P profiles described above in Section~\ref{sec:T-P}. We specify the atmospheric composition using custom equation of state (EOS) files for fixed mixing ratio atmospheres;  for each model, the water mixing ratio is given by the appropriate value from Table~\ref{tab:main_grid}, and the remainder of the atmosphere is composed of hydrogen and helium in solar composition ratios (with the exception of the no-helium model sub-grid, which fills the remainder of the atmosphere with hydrogen only).  For all models, the assumed planetary and stellar radii are $0.238$ $R_{Jup}$ (2.61 $R_{\oplus}$) and $0.216$ $R_\sun$, respectively, corresponding to a GJ 1214b-analog planet.  

The opacity sources included in our \texttt{Exo-Transmit} calculations are those associated with water, molecular hydrogen, and helium.  We include collision-induced opacities for H$_2$-H$_2$ and H$_2$-He pairs; Rayleigh scattering from H$_2$O, H$_2$, and He; and molecular absorption from H$_2$O.  The CIA and Rayleigh scattering opacities are those included in the \texttt{Exo-Transmit} package \citep[see][for details]{kempton17}.  The H$_2$O molecular line opacities are the same ones used in our \texttt{HELIOS} modeling employing our novel prescription for water self-broadening (described in Appendix~\ref{app:water_opac}), but re-binned to a spectral resolution of $R = 1,000$ and reformatted appropriately for \texttt{Exo-Tramsmit}.

As described in Section~\ref{sec:grid}, our baseline model grid consists of clear atmospheres.  For our cloudy model sub-grid (Table~\ref{tab:cloudy_grid}), we \textit{post hoc} place a gray optically thick absorber at the specified cloud-top pressure, effectively truncating the T-P profile and all atmospheric transmission at that location.  This is a highly simplified treatment of the optical properties of true  clouds, but it captures their basic light blocking nature and remains agnostic as to the composition and particle size distribution of the aerosols.

\subsection{Simulated JWST Data and Retrievals \label{sec:retrievals}}

In order to assess the observability of water-rich exoplanet atmospheres with JWST, we generate mock observations of a subset of our model grid using the instrument simulator \texttt{Pandexo} \citep{batalha17c}, and then we retrieve on the simulated spectra with the \texttt{PLATON} retrieval code \citep{zhang19, zhang20}.  The parameters of our simulated observations are selected to align with the benchmark sub-Neptune GJ 1214b, which will be observed in both transmission and thermal emission during JWST's Cycle 1.  The goal of this exercise is not to make specific predictions for GJ 1214b (e.g.\ we do not attempt to match existing observations of its featureless transmission spectrum), but rather to test the performance of retrievals for water-rich sub-Neptunes on a system with realistic physical parameters.

Specifically, we simulate two secondary eclipses with MIRI-LRS ($5-12$ $\mu$m) using our synthetic \texttt{HELIOS} thermal emission spectra.  \citep[A phase curve of GJ 1214b will be obtained in Cycle 1 with MIRI-LRS, bracketed by two secondary eclipses;][]{bean21b}.  For transmission, we  simulate one transit each with NIRCam-Grism+F322W2, NIRCam-Grism+F444W, and MIRI-LRS using our synthetic \texttt{Exo-Transmit} spectra, ultimately combining the data from the two NIRCam modes together when performing a retrieval. (Transits of GJ 1214b will be observed with the NIRCam observing modes during Cycle 1 as part of the Guaranteed Time Observation program, spanning $2.5 - 5$ $\mu$m, and with MIRI-LRS as part of the aforementioned GJ 1214b phase curve.)  The host star is modelled using the stellar SEDs from the PySynPhot package\footnote{\url{ftp://ftp.stsci.edu/cdbs/tarfiles/synphot5.tar.gz}} that is included with \texttt{Pandexo}, interpolated to the parameters of GJ 1214 (parameter values listed in Section~\ref{sec:T-P}) with a J-band magnitude of 9.75 \citep{cutri03}.  As the systematic noise floor level for JWST instruments is not yet known, for our \texttt{Pandexo} calculations we assume a conservative noise floor of 30 ppm based on preliminary estimates \citep{matsuo19, schlawin20, schlawin21}.  All of our \texttt{Pandexo} simulations assume a transit duration of 0.8688 hours, with equivalent amounts of in-transit and out-of-transit observing time.
 
 \begin{figure*}[t]
    \centering
    \includegraphics[width=1.0\textwidth]{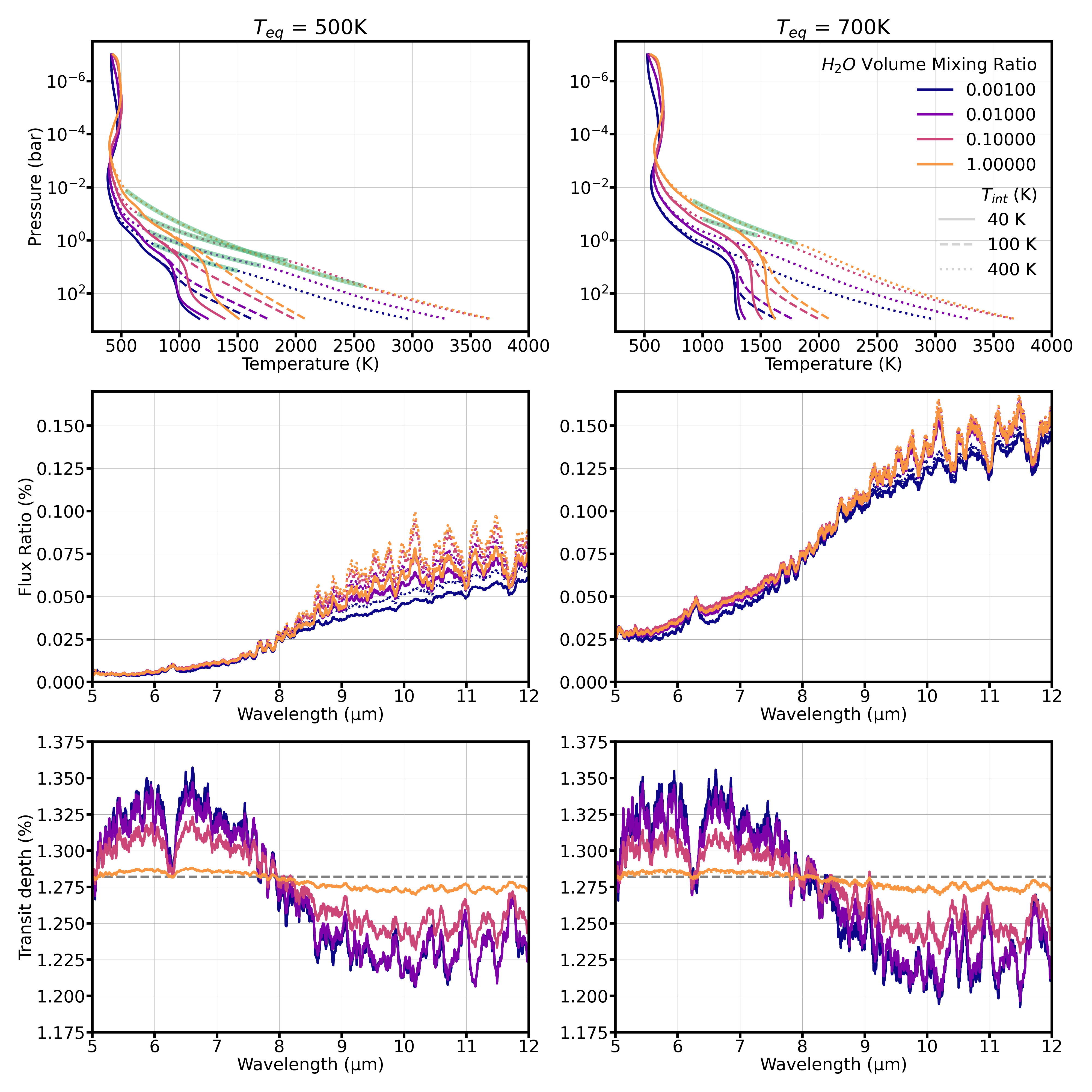}
    \caption{T-P profiles (top panels), emission spectra (middle panels), and transmission spectra (bottom panels) for a planet with $\log(g)=2.9$, orbiting an M-dwarf star.  Models for other values of $\log(g)$ and a Sun-like host star have qualitatively similar behavior.  Left and right panels are for planets with $T_{eq} = 500$ K and $T_{eq} = 700$ K, respectfully.  Colors and line styles are as indicated, denoting models with different water mixing ratios and $T_{int}$ values.  Thick green lines in the upper panels indicate the locations of convection zones.  Dashed gray lines in the lower panels indicate a no-atmosphere planet.  12 combinations of water mixing ratio and $T_{int}$ are plotted in each figure, but in some cases, degeneracies among models result in overlapping (and therefore hidden) lines, as discussed in the text in more detail.  The emission and transmission spectra are plotted for the wavelength range of the JWST / MIRI LRS observing mode, and have been smoothed for better viewing using a running average of 30 points.}
    \label{fig:combined_onerow} 
\end{figure*}

\begin{figure*}
\centering
\includegraphics[width=1.0\textwidth]{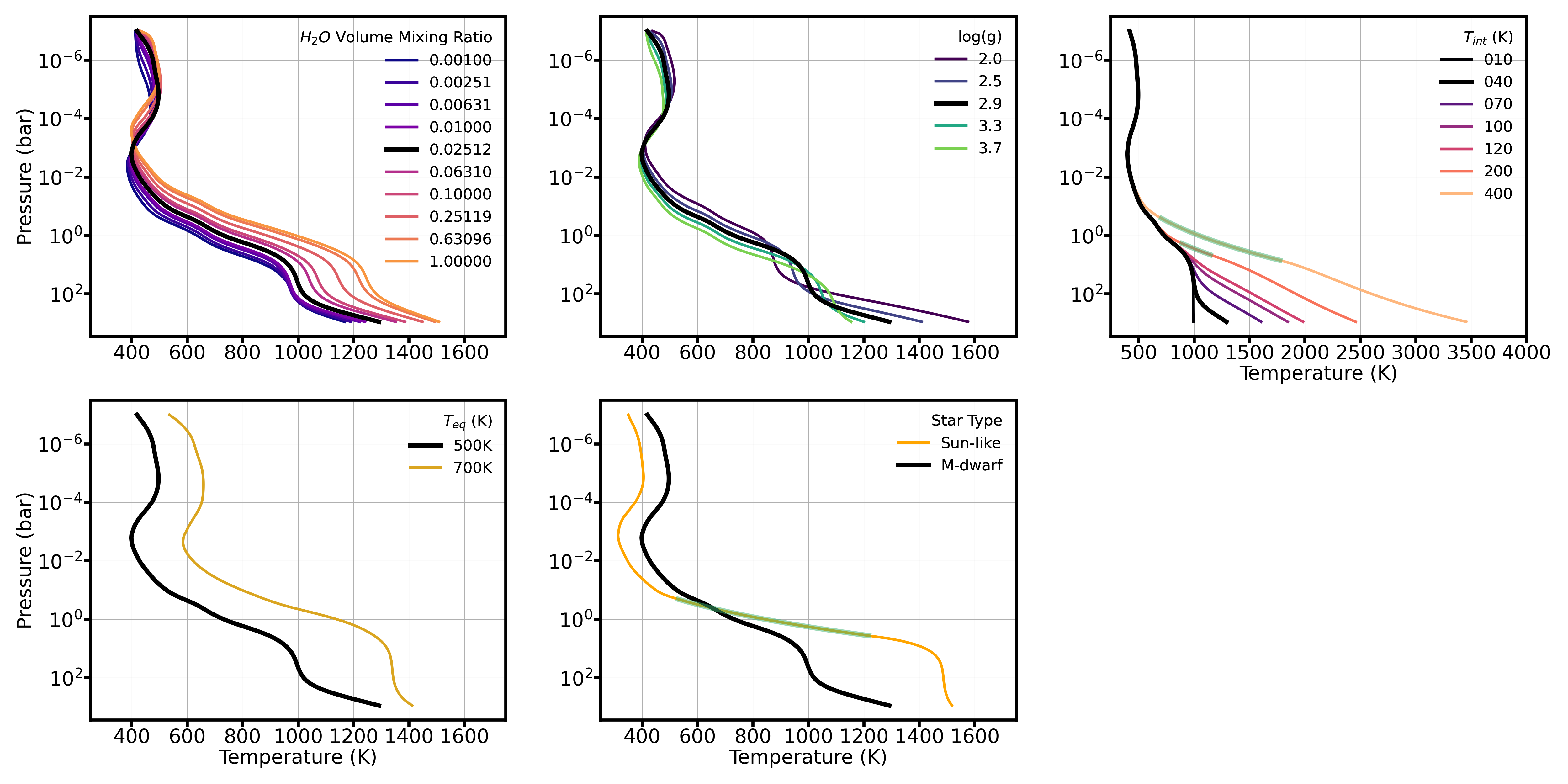}
 \caption{T-P profiles for our baseline model grid, varying one parameter at a time, as indicated.  In each panel, the black line is for a GJ 1214b-like planet with $\log(g) = 2.9$, $T_{int} = 40$ K, $T_{eq} = 500$ K, and an M-dwarf host star, chosen with a volume mixing ratio of 0.02512 --- the rest of the model parameters retain these values, when not explicitly listed in the legend.  Thick green lines denote convection zones.  (Note the different x-axis scale in the upper right-hand panel.)}   
 \label{fig:tp_subplots} 
\end{figure*}

\begin{figure*}
\centering
\includegraphics[width=1.0\textwidth]{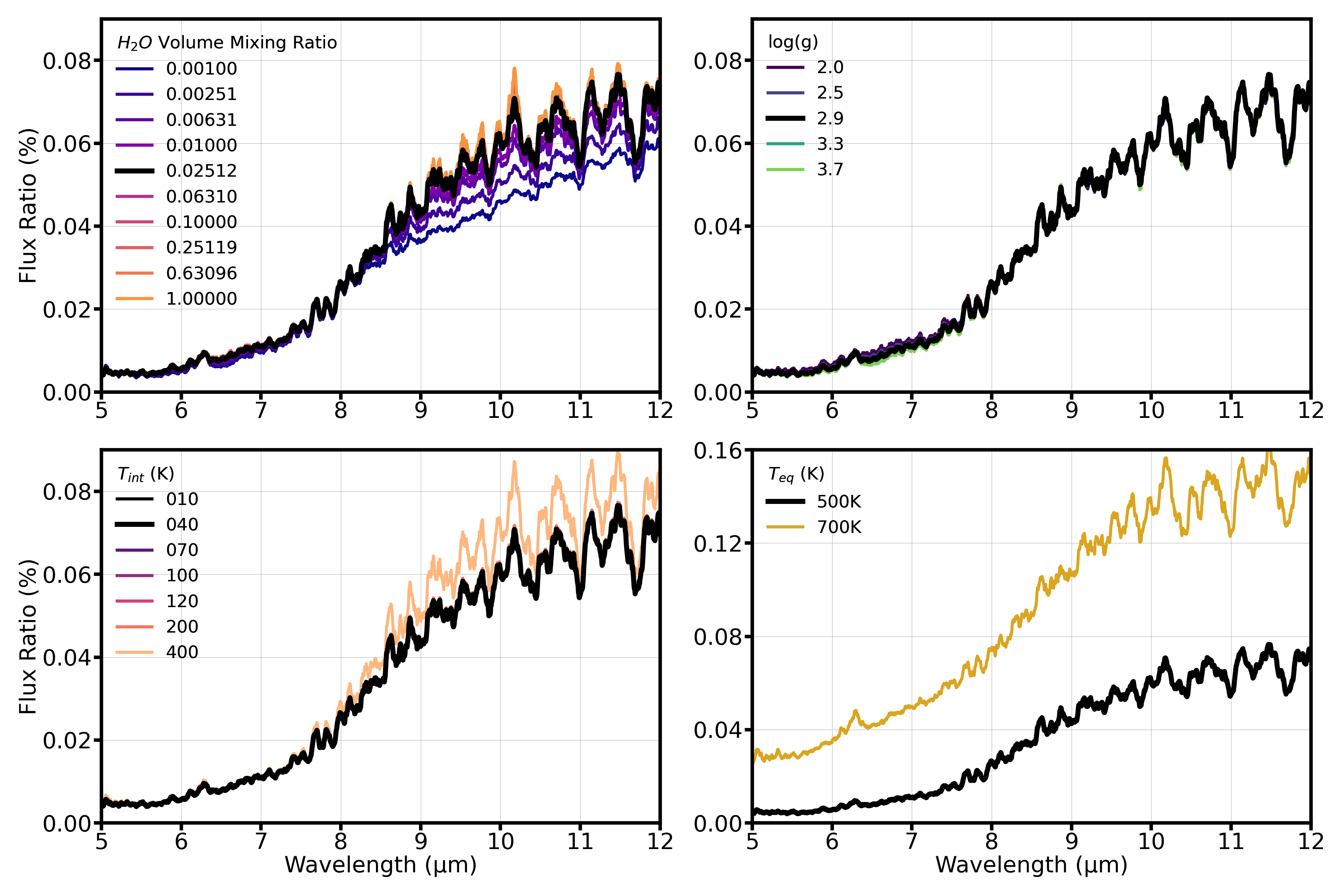}
 \caption{Same as Figure~\ref{fig:tp_subplots}, but for secondary eclipse (emission) spectra.} 
    \label{fig:emiss_subplots} 
\end{figure*}

\begin{figure*}[t!]
\centering
\includegraphics[width=1.0\textwidth]{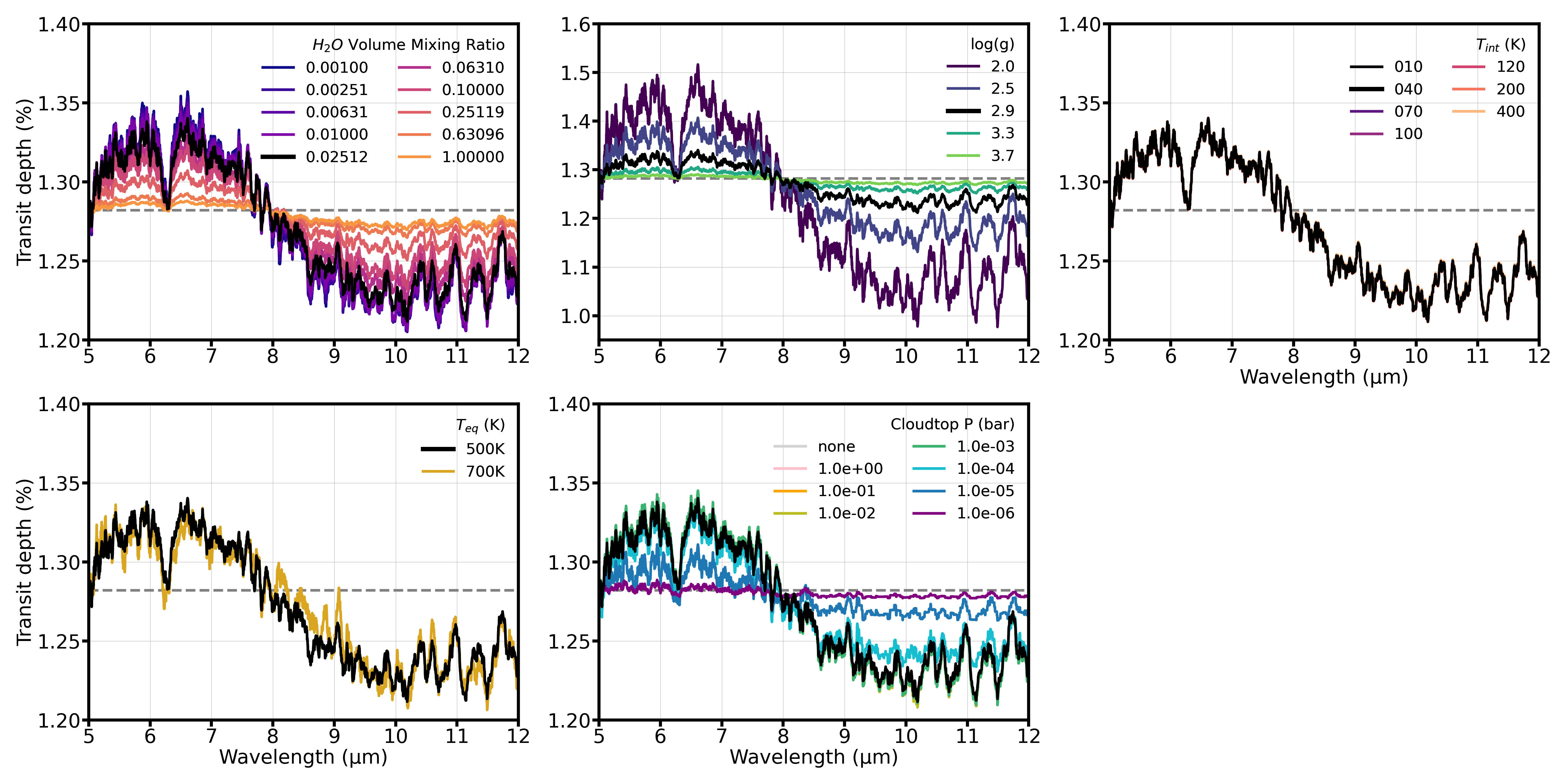}
 \caption{Same as Figure~\ref{fig:tp_subplots}, but for transmission spectra.  The dashed line is for a no-atmosphere case, assuming a planetary radius of  0.238 $R_{Jup}$.  All models have been normalized such that they achieve an equivalent planetary radius when averaged across the entire 5-12 $\mu$m range.  In all but the lower middle panel, the atmosphere is modeled as cloud-free.} 
    \label{fig:trans_subplots} 
\end{figure*}
 
For each of our simulated JWST observations, we run a retrieval using a modified version of the \texttt{PLATON} code \citep{zhang19, zhang20}. Our version of \texttt{PLATON} is set up to perform ``free" retrievals, such that it can recover the abundances of individual species, rather than assuming thermochemical equilibrium.  In our case, we retrieve on a single parameter that sets the atmospheric composition, which is the ratio of water to hydrogen and helium (where the H$_2$ / He ratio is fixed to the solar value of 5.67).  We perform retrievals with our input data binned to spectral resolutions of $R = 20$, 50, and 100.  
For our baseline \texttt{PLATON} runs, we use the water opacities that are provided with the code, which differ from the self-broadened opacities that we used to generate the forward models in our grid.  We purposefully allow for this mismatch in order to assess how well an ``out of the box" retrieval code will work on water-rich atmospheres, and we comment on the outcomes of that experiment in Section~\ref{sec:JWST_results}.  For the transmission retrievals we retrieve on four parameters: the planetary radius, $R_p$, the isothermal atmospheric temperature, $T$, the pressure of a gray cloud-top, $\log (P_{\mathrm{cloud}})$, and the water abundance, $\log (\mathrm{H_2O / (H_2 + He}))$.  For the thermal emission retrievals, we retrieve on 8 total parameters: $R_p$, $\log (P_{\mathrm{cloud}})$, $\log (\mathrm{H_2O / (H_2 + He}))$, and 5 parameters that define the T-P profile following \citet{line13}, $\log (\kappa_{th})$, $\log (\gamma_{1})$, $\log (\gamma_{2})$, $\alpha$, and $\beta$.  Each of the aforementioned parameters is assigned a uniform prior, except for the planetary radius, which is assigned a Gaussian prior with a standard deviation of 0.018 $R_{Jup}$.

\section{Results} \label{sec:results}

\subsection{Baseline Modeling Results}

For a given planet with known stellar host type, instellation, and surface gravity, the two key unknown properties within our model grid are the composition of the planet's atmosphere and its internal heat flux.  We therefore first focus on these two properties.  The top two panels of Figure~\ref{fig:combined_onerow} shows the results of varying composition and $T_{int}$ on our T-P profile modeling for a representative planet with $\log g = 2.9$ orbiting an M-dwarf host star.  We see that larger water abundances result in the atmosphere being optically thick at higher altitude (lower pressure), which in turn elevates the photosphere location and results in a hotter lower atmosphere via a stronger greenhouse effect.  Higher $T_{int}$ heats the atmosphere from below and therefore raises the temperature of the deep atmosphere.  For the highest $T_{int}$ case (400 K), the internal heat contribution can impact the T-P profile to pressures as low as 10 mbar, with potentially observable impacts.  Convective regions (thick green lines in the top panels of Figure~\ref{fig:combined_onerow}) also arise for atmospheres with the highest $T_{int}$, due to strong heating of the photosphere from both above and below in these cases.    

The corresponding emission and transmission spectra are shown in Figure~\ref{fig:combined_onerow} in the middle and bottom panels, respectfully.  Among the emission spectra, high water content atmospheres show stronger spectral features, especially apparent at longer wavelengths.  Interestingly, there is a clear degeneracy among spectra for high water content atmospheres, despite obvious differences in the T-P profiles for these cases.  This arises because, while the photospheric pressure decreases for more water enriched atmospheres (i.e.\ the atmosphere is optically thick higher up), the temperature and temperature gradient around the photosphere remain approximately the same across models. However, for the most water-poor atmospheres, H$_2$-H$_2$ CIA dominates over the H$_2$O opacity redward of 8 $\mu$m. This moves the photosphere higher than it would be with only H$_2$O absorption and, since the temperature decreases with height at this region, leads to a significantly lower emission at these wavelengths.
The cloud-free transmission spectra (bottom panels of Figure~\ref{fig:combined_onerow}) show expected behaviors, with the decreased scale height of the most water-rich atmospheres correspondingly decreasing the amplitude of spectral features.  

As for the impact of internal heating on emission and transmission spectra, as a general rule of thumb, $T_{int}$ only has an observable impact for cases in which its value is approximately equal to or greater than the planet's equilibrium temperature.  Furthermore, the impact of $T_{int}$ on transmission spectra is considerably weaker than on emission spectra because the former probe higher in the atmosphere where stellar heating dominates the energy budget.  These expectations play out in Figure~\ref{fig:combined_onerow}, where only emission spectra with the highest $T_{int}$ values are discernible from the rest of the models; the latter show no sensitivity to $T_{int}$.  The effect of high $T_{int}$ is also much more readily apparent in the lower $T_{eq}$ models. The modeled transmission spectra, which probe higher in the planetary atmosphere, show no dependence on $T_{int}$.  We can therefore conclude that interior heat flux is generally an unimportant parameter for interpreting sub-Neptune spectra, except for the case of thermal emission from planets with very high internal heat fluxes that rival the instellation flux.

Our baseline modeling results generally follow expected trends with respect to the rest of the model input parameters.  We show and describe these here for completeness, and also to highlight several particularly interesting effects. Figures \ref{fig:tp_subplots}, \ref{fig:emiss_subplots}, and \ref{fig:trans_subplots} show the T-P profiles, emission spectra, and transmission spectra respectively as a function of each model input parameter.  We have already discussed in the previous paragraphs how the models respond to changes in water abundance and $T_{int}$.  Higher surface gravities push the photosphere to higher pressures in the atmosphere, which is a straightforward consequence of hydrostatic equilibrium. This results in a less efficient greenhouse effect and lower temperatures at the bottom of the modeled atmosphere, at $P \approx 1000$ bar, which primarily arises because the optically-thick portion of the atmosphere is thinner.  The emission spectra have only a weak dependence on surface gravity, with the most marked differences appearing from 5-8 $\mu$m (and generally in the cores of water absorption bands), which comes about primarily from enhanced pressure broadening at the photosphere for the high surface gravity cases.  The transmission spectra are highly sensitive to surface gravity due to its inverse relation to the atmospheric scale height ($H \propto g^{-1}$).  Similarly, the atmospheric scale height ($H \propto T$) is responsible for the temperature dependence of the transmission spectra, seen in the lower left-hand panel of Figure~\ref{fig:trans_subplots}.

Temperature-pressure profiles for the Sun-like host star differ quite significantly from equivalent models for the M-dwarf star (Figure~\ref{fig:tp_subplots}, middle right-hand panel).  This is a result of the stellar energy being deposited at different layers of the atmosphere due to the interaction between the opacity in the planet's atmosphere relative to the wavelengths at which the star emits most of its energy.  For the M-dwarf host star, because it emits mostly at longer wavelengths where water water vapor also absorbs efficiently, there is increased upper atmosphere heating (leading to a moderate thermal inversion, but also hotter upper-atmosphere temperatures overall), and the lower atmosphere and surface are consequently cooler to maintain energy balance.  For the Sun-like star, the incoming (mostly visible light) radiation from the star penetrates much more deeply into the atmosphere.  In combination with the planet's outgoing thermal emission, this produces strong heating in the middle atmosphere around pressures of 1 bar, driving the atmosphere into a convective regime.  We do not plot transmission or secondary eclipse spectra for the Sun-like host star because the observational signals would be too small to be measured with JWST.  

Finally, the addition of a cloud-top in the transmission models (Figure~\ref{fig:trans_subplots}, lower middle panel) alters the shape of the spectra in the expected way.  Higher altitude (i.e.\ lower pressure) clouds mute and ultimately erase spectral features from the gaseous atmosphere.  The clouds don't significantly impact the transmission spectra unless they are optically thick at pressures of $\lesssim$1 mbar, above the location of the transmission photosphere for the gaseous atmosphere.  There is a well-known degeneracy between how clouds and high mean molecular weight shape transmission spectra, which can be seen by comparing the upper left and middle right panels of Figure~\ref{fig:trans_subplots}.  We explore this in more detail in Section~\ref{sec:degeneracy}.

\subsection{Degeneracy Between Model Parameters \label{sec:degeneracy}}

\begin{figure*}
\plotone{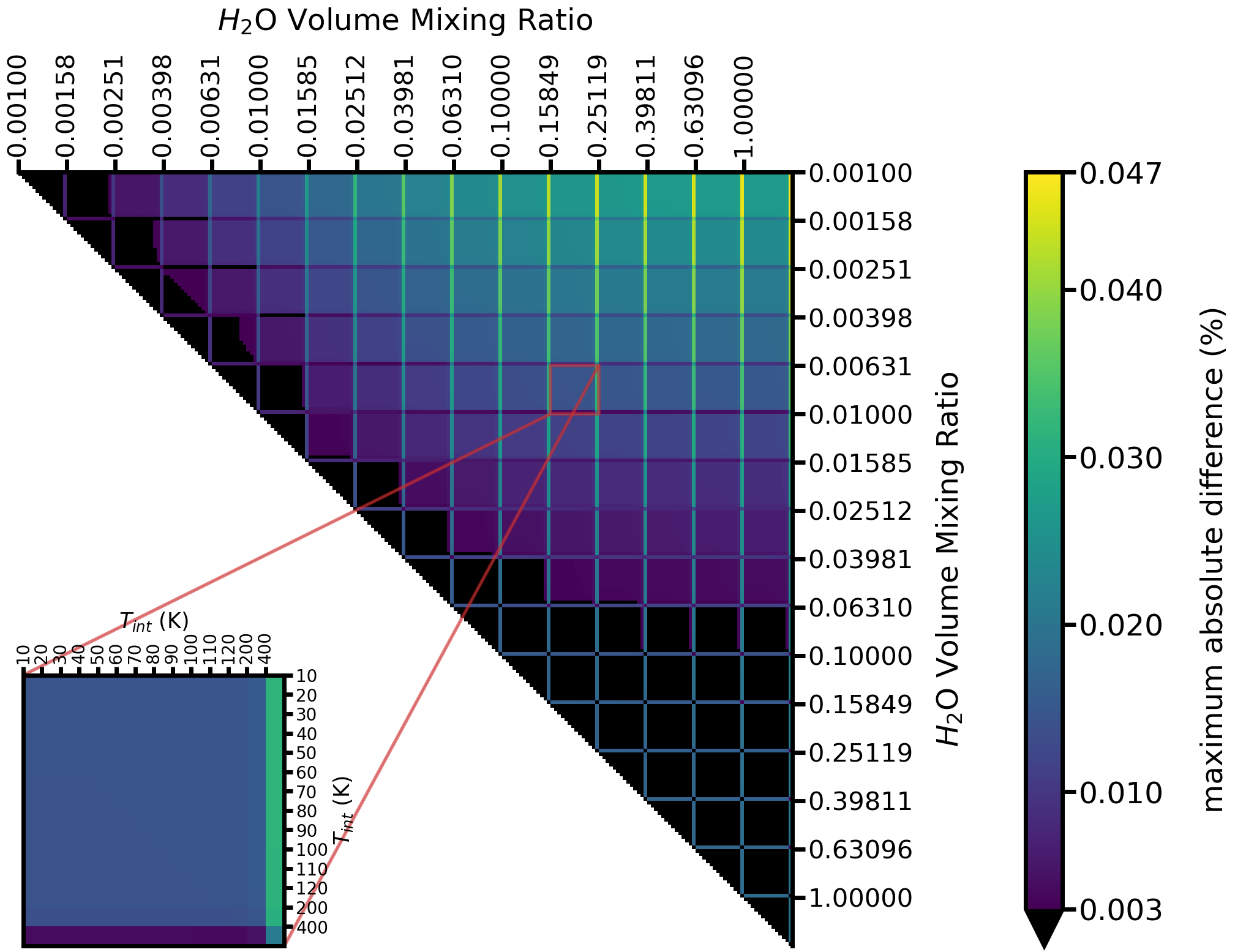}
 \caption{Maximum absolute difference between pairs of emission spectra for the case of a planet with $\log(g) = 2.9$, \mbox{$R_p = 0.238$ $R_{Jup}$}, $R_* = 0.216$ $R_{\odot}$, and $T_{eq} = 500$~K.  The maximum absolute difference is plotted in units of percent (color scale), as a function of H$_2$O volume mixing ratio (large grid cells) and $T_{int}$ (smaller embedded grid cells --- see zoom-in panel at lower left).  Regions shaded in black are for pairs of spectra with differences of no more than 30 ppm, corresponding to a conservative noise floor for JWST-MIRI.  From this figure we can conclude that $T_{int}$ only impacts emission spectra at an observable level for values in excess of 200 K and that low water mixing ratios are more readily distinguishable than higher values.  This plot is qualitatively similar for other planetary and stellar parameters (e.g.\ $R_p$, $R_{\star}$, $g$, $T_{eq}$), but the magnitude of the effect changes, and the black-shaded region grows / shrinks accordingly.} 
    \label{fig:emiss_megaplot} 
\end{figure*}

\begin{figure*}
\plotone{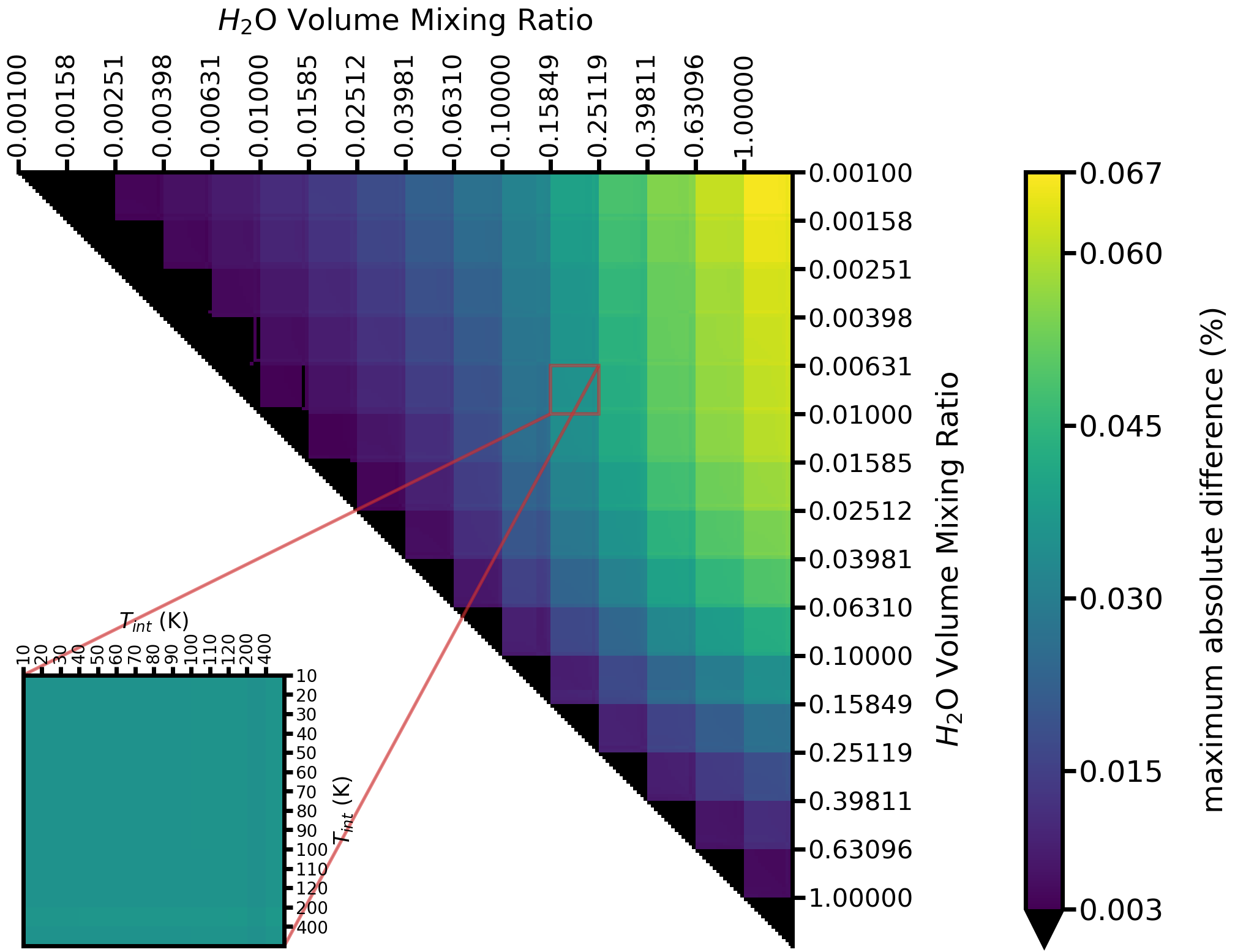}
 \caption{Same as Figure~\ref{fig:emiss_megaplot}, but for cloud-free transmission spectra.  Prior to calculating the maximum absolute difference, we normalize all spectra to provide an identical average transit depth (corresponding to the no-atmosphere case) across the MIRI-LRS bandpass.  From this figure we can conclude that $T_{int}$ has a negligible impact on transmission spectra at an observable level and that high water mixing ratios are more readily distinguishable than lower values.} 
    \label{fig:trans_megaplot} 
\end{figure*}

\begin{figure*}
\plotone{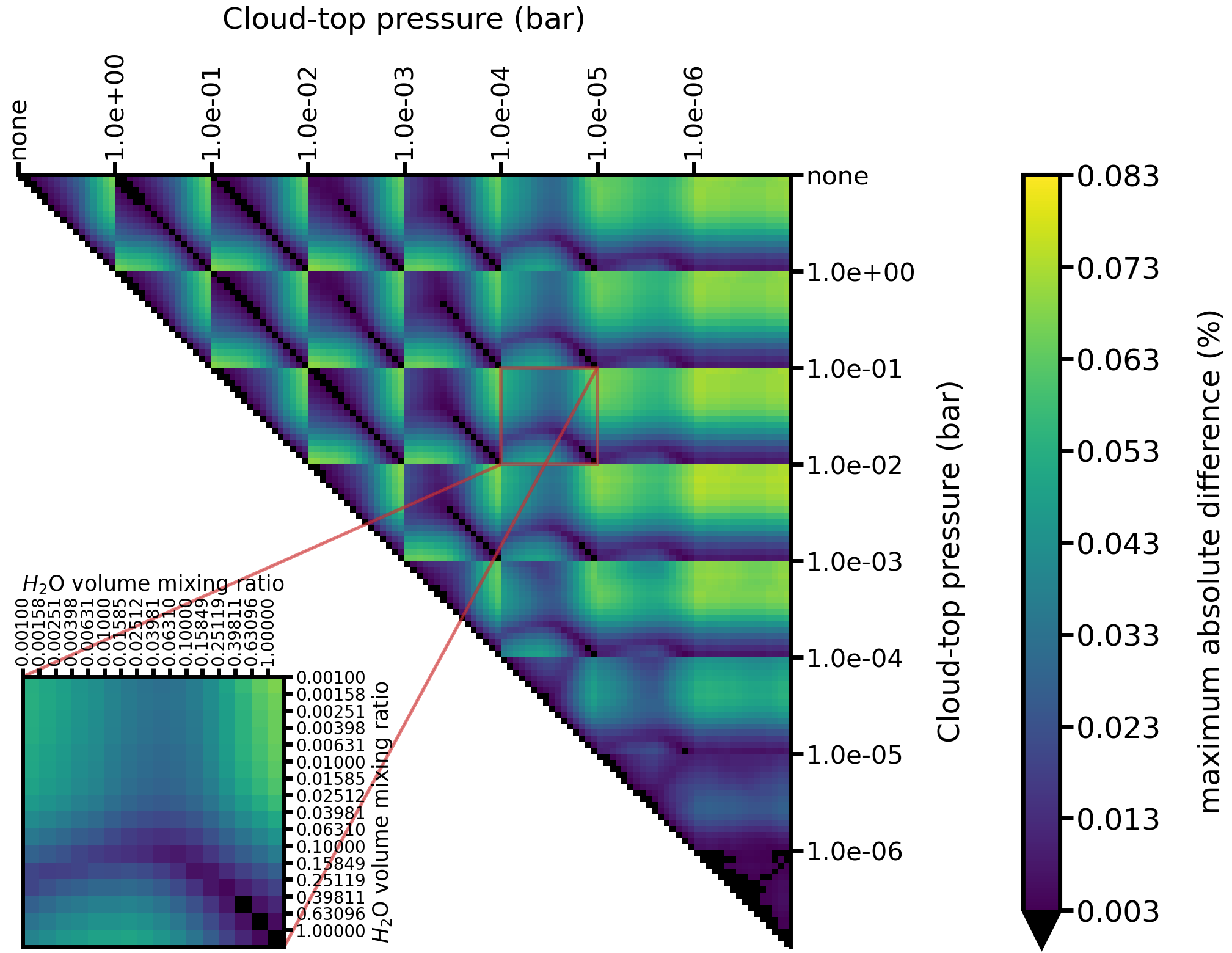}
 \caption{Similar to Figure~\ref{fig:trans_megaplot}, but here $T_{int}$ is held constant at a value of 40 K, and cloud-top pressure (large grid cells) and water abundance (smaller embedded grid cells --- see zoom-in panel at lower left) are varied as the two parameters of interest.  In the absence of impact of clouds, the large grid cells should look identical to Figure~\ref{fig:trans_megaplot}, mirrored over the upper-left to lower-right diagonal (e.g.\ see the upper left-hand large grid cells).  The presence of low-pressure (high-altitude) clouds alter the degree of degeneracy between spectra, as detailed in the text.} 
    \label{fig:cloudy_megaplot} 
\end{figure*}

We now assess the degree of degeneracy between the spectra produced in our model grid to understand which parameters future data sets might be able to constrain.  To quantify degeneracy, we use a metric that is the maximum difference between two data points at the same wavelength in a given pair of spectra: $\mathrm{max}(|S1_i - S2_i|)$, where $S1$ and $S2$ are the two spectra being compared, and $i$ is the index over wavelength.  This metric is useful because it distills the entirety of a pair of spectra down to a single value.  As a result, however, it is unable to convey any useful information about the wavelength-dependent differences between spectra.  Therefore, for single-planet studies or observational data sets a (much more computationally intensive) retrieval analysis will be more informative to assess the degree of degeneracy between parameters.  

We plot our degeneracy metric for the emission spectrum grid in Figure~\ref{fig:emiss_megaplot}.  Prior to calculating the degeneracy metric, each spectrum in the grid is reduced down to the wavelength range and resolution of the JWST MIRI-LRS observing mode ($5-12$ $\mu$m and $R=100$).  We focus on MIRI-LRS because this is the observing mode for thermal emission measurements with JWST that maximizes the secondary eclipse depth.  

Two key results arise from Figure~\ref{fig:emiss_megaplot}.  The first is that, as already discussed, emission spectra are mostly insensitive to the internal heat flux unless it is very large ($T_{int} \gtrsim 200$ K for the case plotted), and therefore it is typically not possible to distinguish between planets with different $T_{int}$ values but otherwise the same planetary parameters.  Secondly, emission spectra with lower water abundance are more readily distinguishable from one another than those with higher water concentrations.  This can be seen by the blue-green regions in Figure~\ref{fig:emiss_megaplot}, indicating a lower degree of degeneracy, lining up across the top edge of the plot.  For the planetary scenario plotted in Figure~\ref{fig:emiss_megaplot}, atmospheres with water concentrations greater than $\sim$10\% are unlikely to be distinguishable with JWST thermal emission observations because the \emph{maximum} difference between any two spectra is less than 30 ppm.  

Transmission spectra display different behavior. Shown in Figure~\ref{fig:trans_megaplot} is our degeneracy metric for the grid of cloud-free transmission spectra, also considering the wavelength range and resolution of the MIRI-LRS JWST observing mode.  For the planetary parameters plotted ($g = 800$ cm s$^{-2}$, $T_{eq} = 500$ K, and GJ 1214b-like planetary and stellar radii), transmission spectra are less degenerate overall compared to thermal emission spectra due to higher signal-to-noise in transmission.  Transmission spectra are even less sensitive to the selected value of $T_{int}$ because transmission spectroscopy probes higher locations in a planet's atmosphere where the internal heat flux is even less likely to impact the local thermal profile.  Contrary to the thermal emission spectra, cloud-free transmission spectra are more degenerate at \emph{low} water abundance than at high water abundance, which can be seen from the yellow-green regions lining up across the right-hand edge of Figure~\ref{fig:trans_megaplot}.  However, for the idealized case examined in Figure~\ref{fig:trans_megaplot}, almost all abundance pairings can in principle be distinguished from one another at an observable level, given that spectral differences manifest at levels well above the anticipated noise floors of JWST instruments.  

It is important to consider the potential presence of aerosols in exoplanet atmospheres, because such particles are known to mute transmission spectral features.  In Figure~\ref{fig:cloudy_megaplot}, we re-plot the results from Figure~\ref{fig:trans_megaplot} for an internal temperature of $T_{int} = 40$ K but also including an optically thick cloud deck at varying pressures as an additional parameter.  If clouds did not introduce an additional source of degeneracy, the large grid cells in Figure~\ref{fig:cloudy_megaplot} would identically match Figure~\ref{fig:trans_megaplot}, mirrored over the diagonal running from upper left to lower right.  The smearing of this (Figure~\ref{fig:trans_megaplot}-like) pattern in the large grid cells toward the right and toward the bottom of the plot is indicative of the growing importance of clouds on the transmission spectrum as the cloud-top pressure decreases.  Deep purple regions along the right-hand side of Figure~\ref{fig:cloudy_megaplot} reveal the well-known degeneracy between cloudy and high mean molecular weight atmospheres.  Therefore, while transmission spectra are less degenerate for high water content atmospheres in the absence of clouds, once the possibility of aerosols is considered that statement becomes less definitive.  Still, the near absence of black regions on Figure~\ref{fig:cloudy_megaplot} indicates that the cloud / composition degeneracy can be broken for sufficiently high signal-to-noise spectra.  

In total, the degeneracy analysis shown in Figures~\ref{fig:emiss_megaplot} -- \ref{fig:cloudy_megaplot} reveals the complementary nature of transmission and thermal emission measurements in uniquely constraining atmospheric water abundances, for cases in which there is no prior knowledge about the water concentration.  Emission spectroscopy is more sensitive to low water abundance, whereas transmission spectroscopy is more sensitive to high water abundance, and degeneracies with clouds can potentially be broken by observing an atmosphere with more than one technique.

\subsection{Assessing the Impact of Novel Treatments of Water-Rich Atmospheric Composition \label{sec:treatment_results}}

\begin{figure}[ht]
    \centering
    \includegraphics[width=0.45\textwidth]{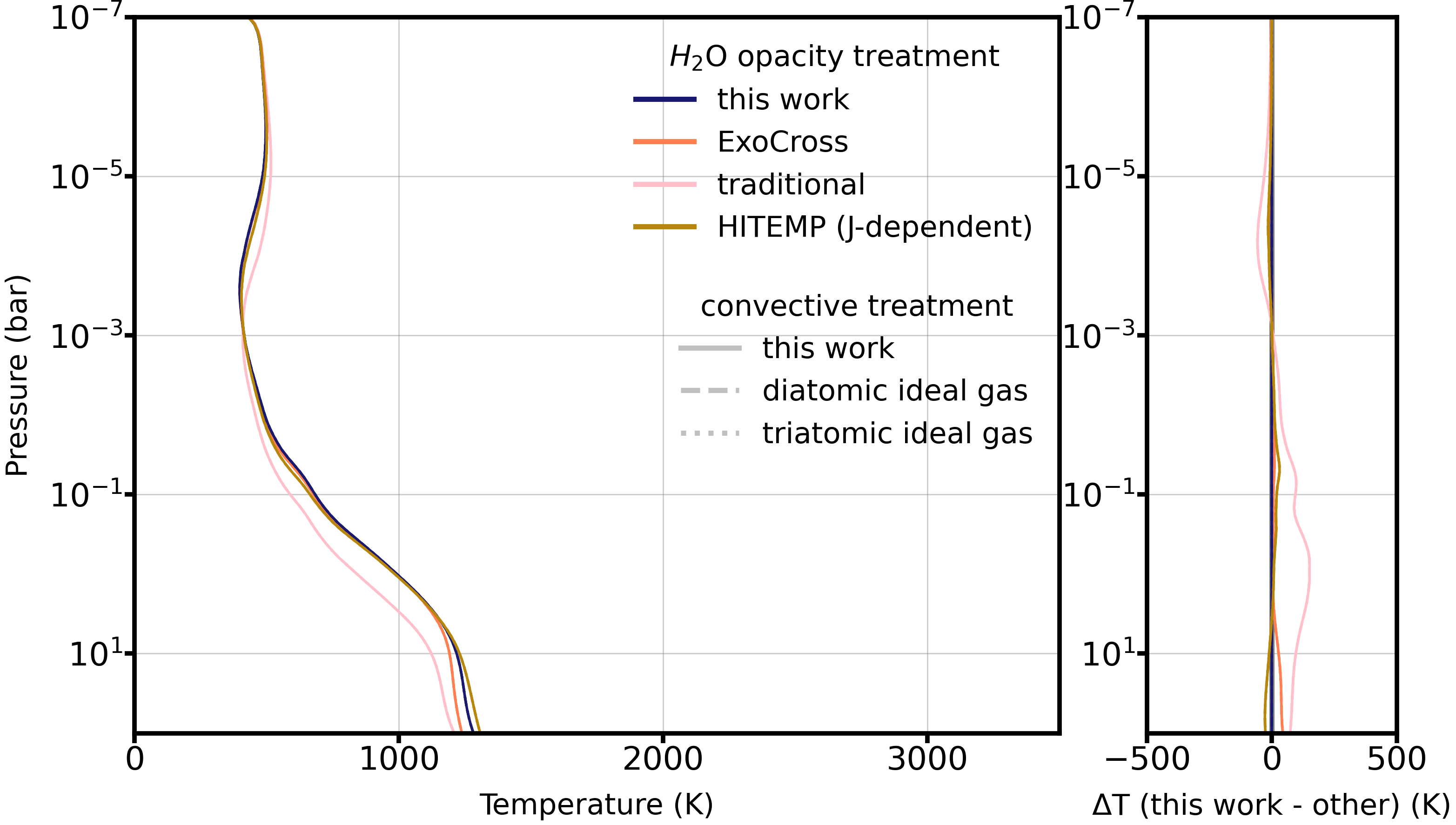}
    \caption{Left: T-P profiles for a planet with $T_{int} = 40$ K, $\log g = 2.9$, an M-dwarf host star, and a 100\% water vapor atmosphere. The various curves are for otherwise equivalent models run with the different treatments of pressure broadening and convective lapse rates listed in Table~\ref{tab:treatment_grid} and indicated in the figure legend.  Right: Temperature differences for each model, relative to the modeling approach used in this work (solid dark-blue line).  None of the models depicted here form convective regions, so the dashed lines directly overlap the solid lines.} 
    \label{fig:treat_tp_onerow_Tint040_logg2.9} 
\end{figure}

\begin{figure}[ht]
    \centering
    \includegraphics[width=0.45\textwidth]{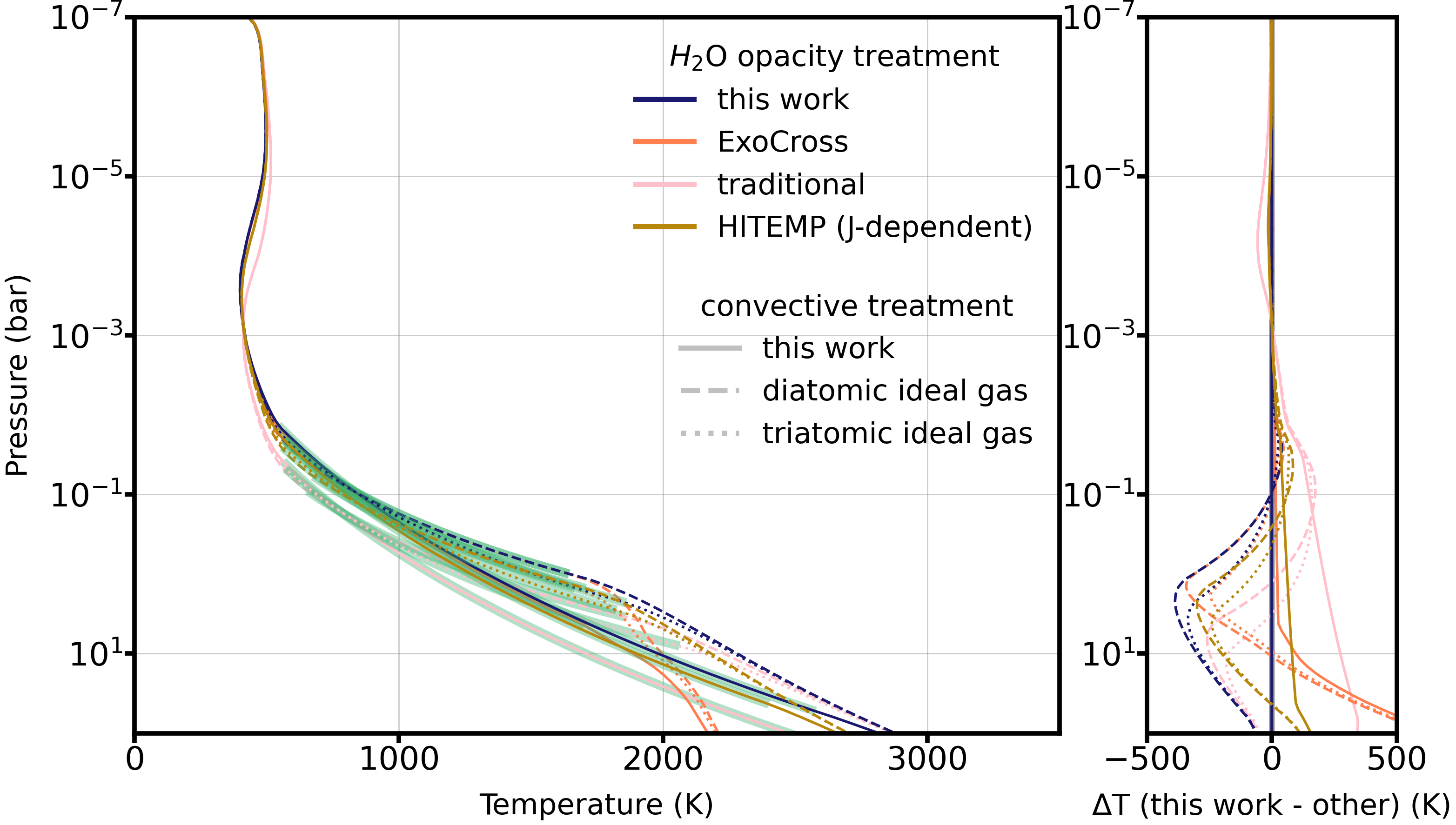}
    \caption{Same as Figure~\ref{fig:treat_tp_onerow_Tint040_logg2.9}, except for a planet with $T_{int} = 400$ K.  Here convective regions do form in the lower atmosphere, which are indicated in the left panel with thick green lines.} 
    \label{fig:treat_tp_onerow_Tint400_logg2.9} 
\end{figure}

\begin{figure}[ht]
    \centering
    \includegraphics[width=0.45\textwidth]{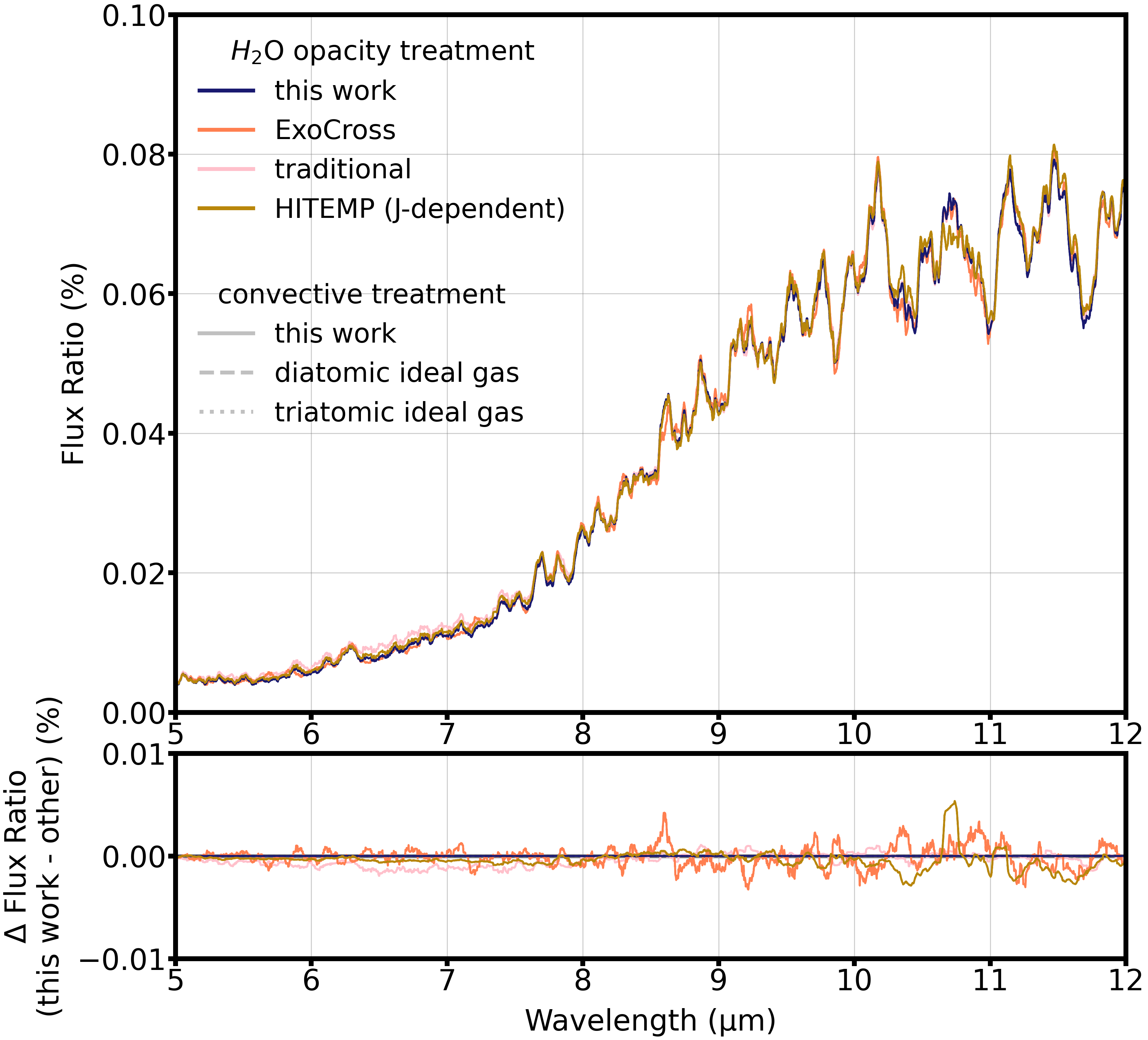}
    \caption{Top: Secondary eclipse spectra  over the MIRI-LRS wavelength range resulting from the T-P profiles in Figure~\ref{fig:treat_tp_onerow_Tint040_logg2.9} for a planet with $T_{int} = 40$ K.  Bottom: Flux ratio differences between each spectrum, relative to the modeling approach used in this work (solid dark-blue line).} 
    \label{fig:treat_emiss_onerow_Tint040_logg2.9} 
\end{figure}

\begin{figure}[ht]
    \centering
    \includegraphics[width=0.45\textwidth]{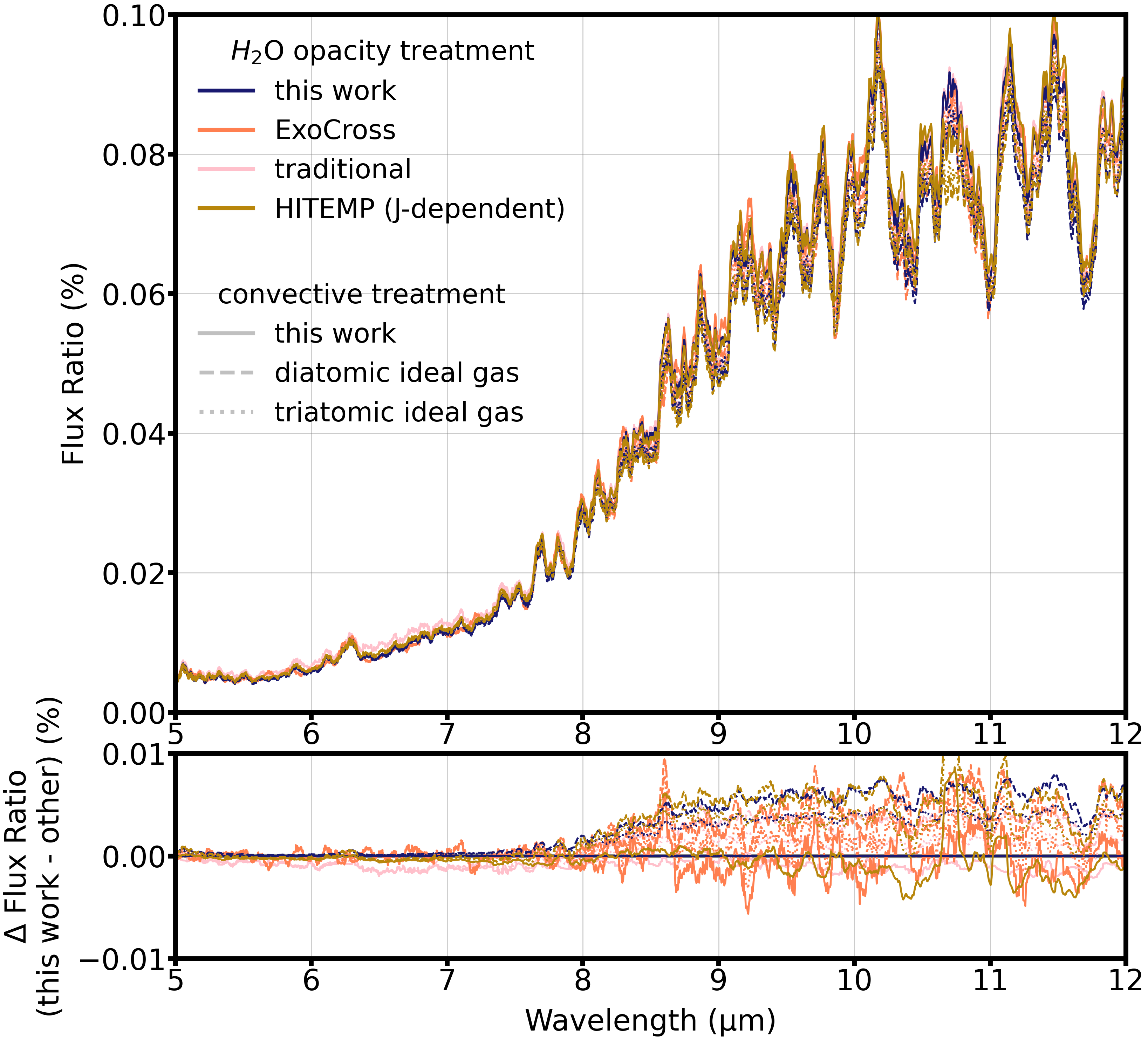}
    \caption{Same as Figure~\ref{fig:treat_emiss_onerow_Tint040_logg2.9}, but using the T-P profiles in Figure~\ref{fig:treat_tp_onerow_Tint400_logg2.9} for a planet with $T_{int} = 400$ K.} 
    \label{fig:treat_emiss_onerow_Tint400_logg2.9} 
\end{figure}

\begin{figure}[ht]
    \centering
    \includegraphics[width=0.45\textwidth]{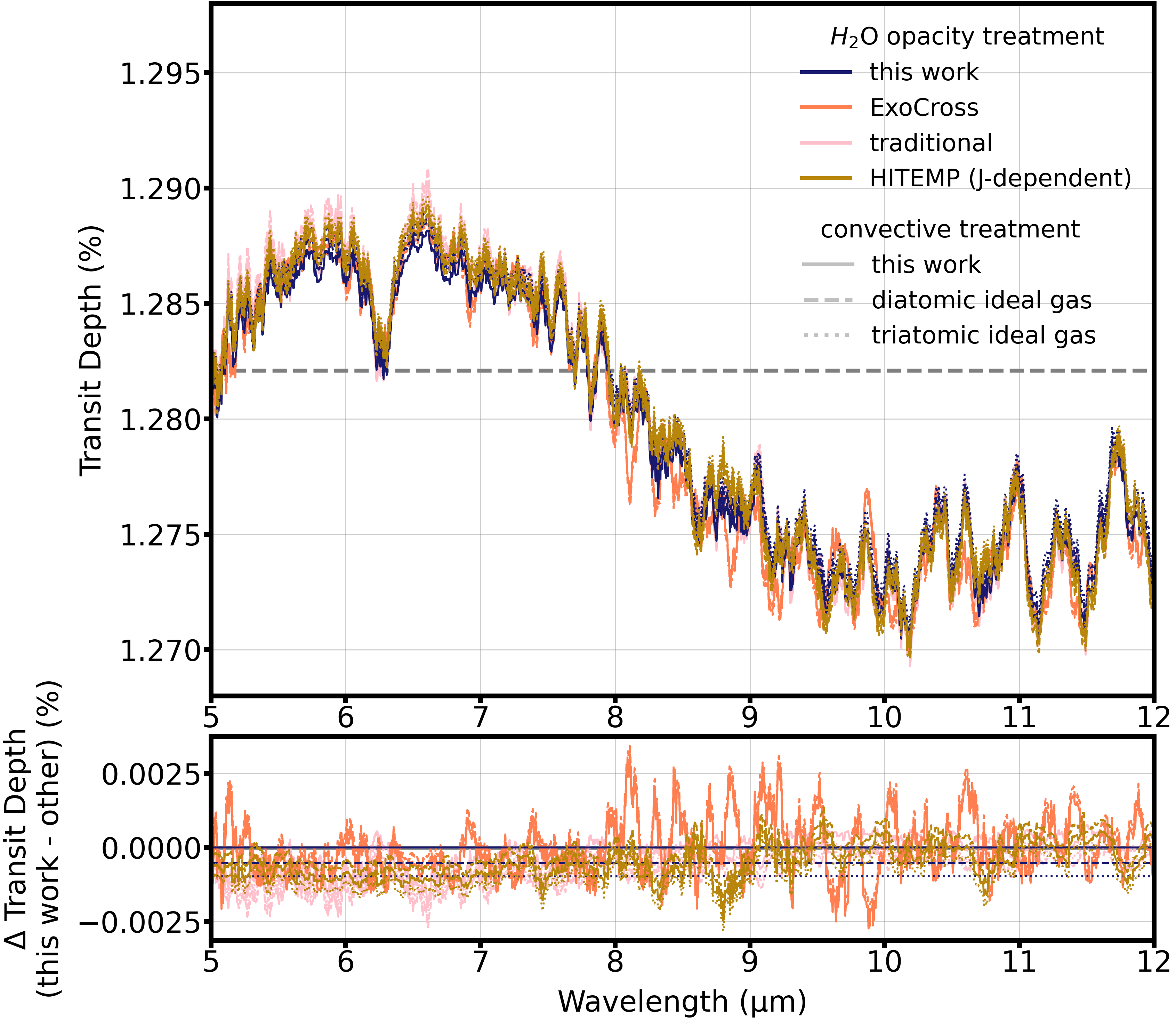}
    \caption{Top: Transmission spectra  over the MIRI-LRS wavelength range resulting from the T-P profiles in Figure~\ref{fig:treat_tp_onerow_Tint400_logg2.9} for a planet with $T_{int} = 400$ K.  Bottom: Transit depth differences between each spectrum, relative to the modeling approach used in this work (solid dark-blue line).  The very slight offset among models with different convective lapse rate treatments (solid black vs.\ dashed black lines) arises because the atmospheric temperature differs slightly between those two cases, giving rise to a small difference in scale height.} 
    \label{fig:treat_trans_onerow_Tint400_logg2.9} 
\end{figure}

As described in Section~\ref{sec:methods}, our modeling techniques include novel treatments of self-broadened water opacities (Appendix~\ref{app:water_opac}) and convective lapse rates (Appendix~\ref{app:adiabatic}).  Here we assess the impact of these novel treatments against a set of comparison models that use H$_2$-broadened molecular line opacities and the assumption of an ideal diatomic or triatomic gas for calculating adiabatic lapse rates, which are standard practices in many exoplanet models.  We aim here to determine the degree to which our updated more self-consistent treatments affect the modeling outcomes.  We focus our following analysis on an atmosphere composed entirely of water vapor to maximize the impact of our novel H$_2$O treatments.  

Figures~\ref{fig:treat_tp_onerow_Tint040_logg2.9} and \ref{fig:treat_tp_onerow_Tint400_logg2.9} show the T-P profiles that result from models using the three different sets of H$_2$O opacities and three approaches for calculating convective lapse rates described at the end of Section~\ref{sec:grid}.  Several key results emerge.  First of all, both sets of self-broadened H$_2$O opacities \citep[those from this work and from][]{gharib19} produce T-P profiles that agree reasonably well with one another but that differ quite significantly, by up to a few hundred Kelvin, from those calculated with the ``traditional'' H$_2$-broadened opacities.  Secondly, the non-ideal gas calculation of the adiabatic lapse rates for water also impacts the T-P profiles at the level of hundreds of Kelvins, although this result is only applicable to atmospheres that develop convective zones, which primarily occurs for the highest $T_{int}$ values explored in this work.  Finally, we find that the most marked departures between the traditional modeling approach (dashed and dotted pink lines in Figures~\ref{fig:treat_tp_onerow_Tint040_logg2.9} and \ref{fig:treat_tp_onerow_Tint400_logg2.9}) and the updated approaches of this work (solid dark blue lines) tend to occur at depth.  This is because pressure broadening and departures from ideal gas behavior both become increasingly important at high pressures.  We therefore expect our updated treatments of opacities and lapse rates to have minimal impact on spectroscopic observables (as we will see below) but more considerable implications for models of the interior structure of water-rich planets that rely on accurate calculations of atmospheric temperatures at depth.  

Figures~\ref{fig:treat_emiss_onerow_Tint040_logg2.9} and \ref{fig:treat_emiss_onerow_Tint400_logg2.9} show the thermal emission spectra that result from the T-P profiles in Figures~\ref{fig:treat_tp_onerow_Tint040_logg2.9} and \ref{fig:treat_tp_onerow_Tint400_logg2.9}, respectively.  Differences between the various treatments of line broadening and lapse rates are most apparent for the high $T_{int}$ case (Figure~\ref{fig:treat_tp_onerow_Tint400_logg2.9}). This is primarily caused by differences in the underlying temperature profiles at pressures of $0.1-1$ bar for the high $T_{int}$ models.  For a GJ~1214b-analog planet, the largest secondary eclipse variations for the different model treatments occur at the $\sim$100 ppm level, which is potentially observable with JWST.  However, for lower values of $T_{int}$, which are more likely for the evolved sub-Neptunes that will be observed with JWST, the secondary eclipse depths only vary by a few tens of ppm at most.  We conclude that atmosphere models predicting secondary eclipse spectra of sub-Neptunes for JWST can mostly get by with ``standard" modeling treatments such as the ideal gas approximation and H$_2$-only pressure broadening, unless the planet in question is suspected to have an especially high degree of internal heating.

Figure~\ref{fig:treat_trans_onerow_Tint400_logg2.9} shows the transmission spectra that result from the T-P profiles in Figure~\ref{fig:treat_tp_onerow_Tint400_logg2.9}, for a planet with $T_{int} = 400$~K.  Because transmission probes much higher in the atmosphere than emission spectroscopy, and pressure broadening and convection are both less important at lower pressures, the differences between our model treatments are far less apparent for transmission spectra.  The maximum differences between model treatments for a GJ 1214b-like planet are $\sim$25 ppm --- just on the verge of having an observable impact, depending on actual JWST noise floor values.  The differences between transmission spectra in Figure~\ref{fig:treat_trans_onerow_Tint400_logg2.9} are primarily driven by the different input water line lists used in \citet{gharib19} vs.\ this work, rather than the pressure broadening or convection treatments.  As with our secondary eclipse modeling, we conclude that transmission spectra are also not strongly impacted at an observable level by the details of the pressure broadening or convective model treatments, but it is still critical to make use of accurate line lists.  

\subsection{JWST Simulation Results \label{sec:JWST_results}}

\begin{figure*}[ht]
    \centering
    \includegraphics[width=0.9\textwidth]{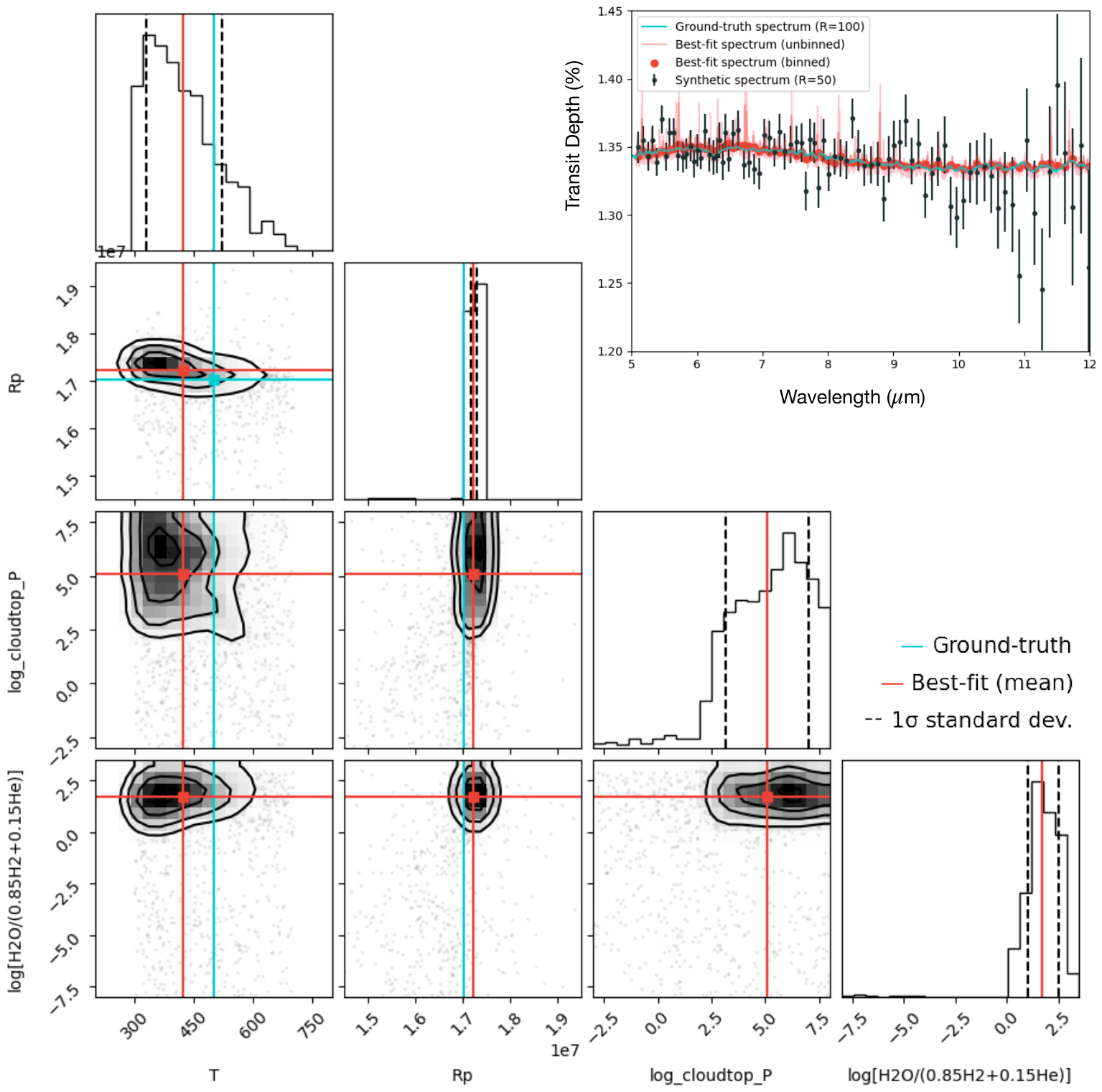}
    \caption{MIRI transmission retrieval for an input model with 100\% H$_2$O and no cloud.  The input spectrum is binned to a resolution of $R = 50$ prior to running the retrieval.  All quantities are given in SI units (i.e.\ pressure in Pa, $R_p$ in meters), and retrieved quantities are defined in Section~\ref{sec:retrievals}.  The best-fit spectrum is plotted in the inset to the upper right.} 
    \label{fig:retrieval_trans_high_H2O} 
\end{figure*}

\begin{figure*}[ht]
    \centering
    \includegraphics[width=0.9\textwidth]{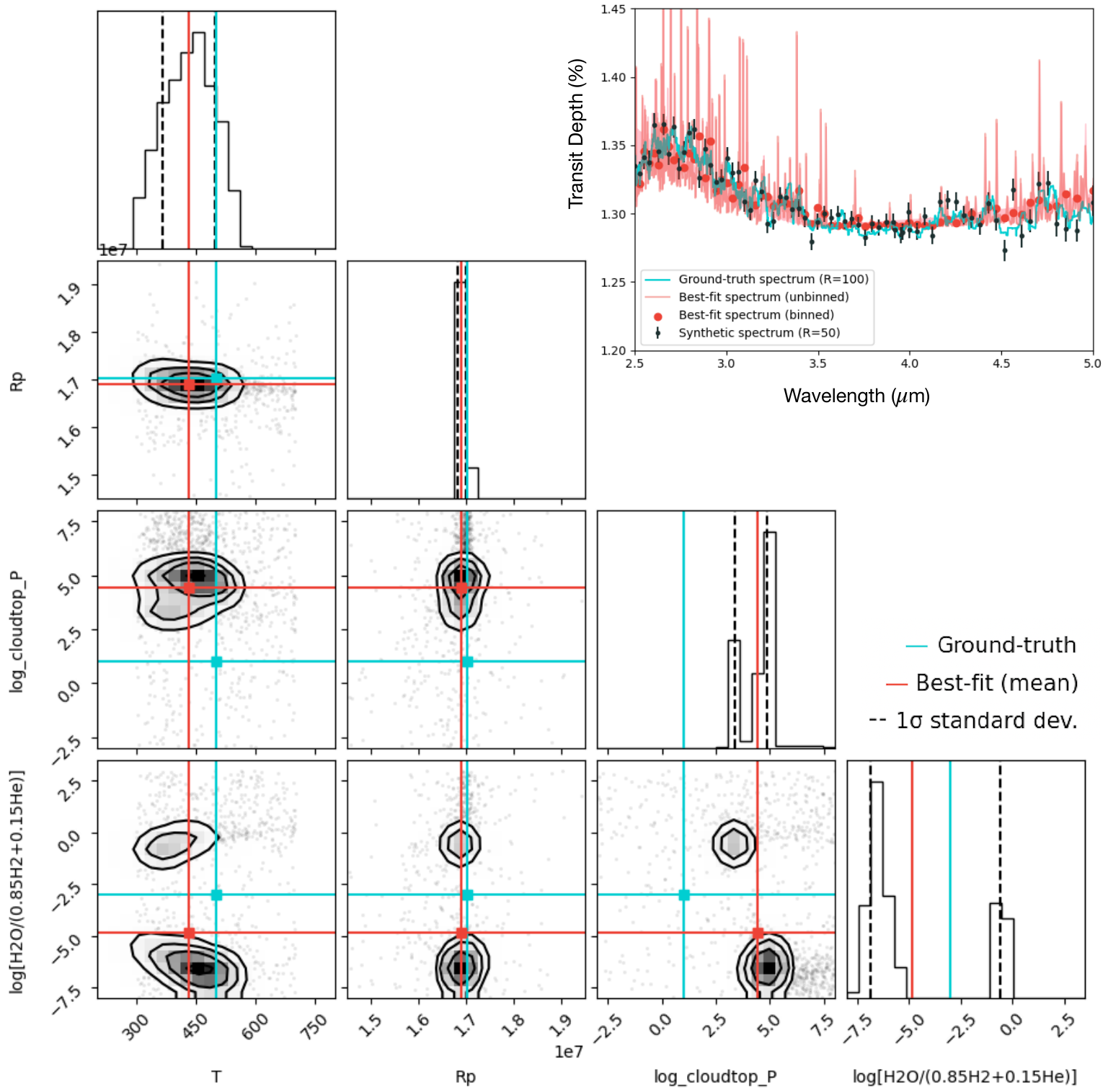}
    \caption{NIRCam transmission retrieval for an input model with 0.1\% H$_2$O and a cloud at 0.1 mbar.  The input spectrum is binned to a resolution of $R = 50$ prior to running the retrieval.  The water abundance and cloud-top pressure are not retrieved accurately, which is due to subtle mismatches between the input model and the retrieval code's forward model.} 
    \label{fig:retrieval_trans_low_H2O} 
\end{figure*}

\begin{figure*}[ht]
    \centering
    \includegraphics[width=0.98\textwidth]{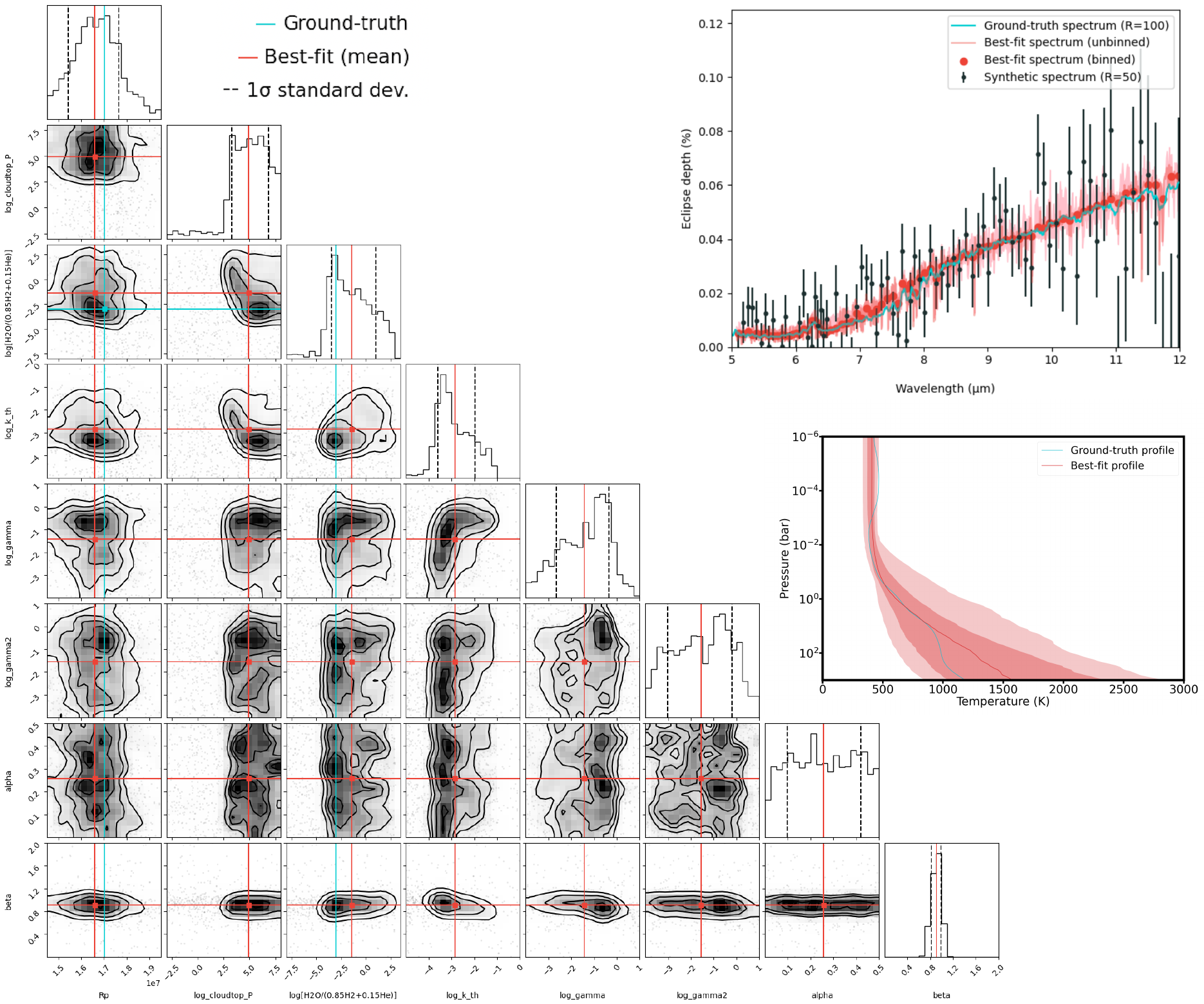}
    \caption{MIRI thermal emission retrieval for an input model with 0.1\% H$_2$O.  The input spectrum is binned to a resolution of $R = 50$ prior to running the retrieval.  The upper shows the best-fit emission spectrum.  The lower inset shows the best-fit T-P profile with 1-$\sigma$ and 2-$\sigma$ errors in dark red and light red, respectively.} 
    \label{fig:retrieval_emiss_low_H2O} 
\end{figure*}

\begin{figure*}[ht]
    \centering
    \includegraphics[width=0.98\textwidth]{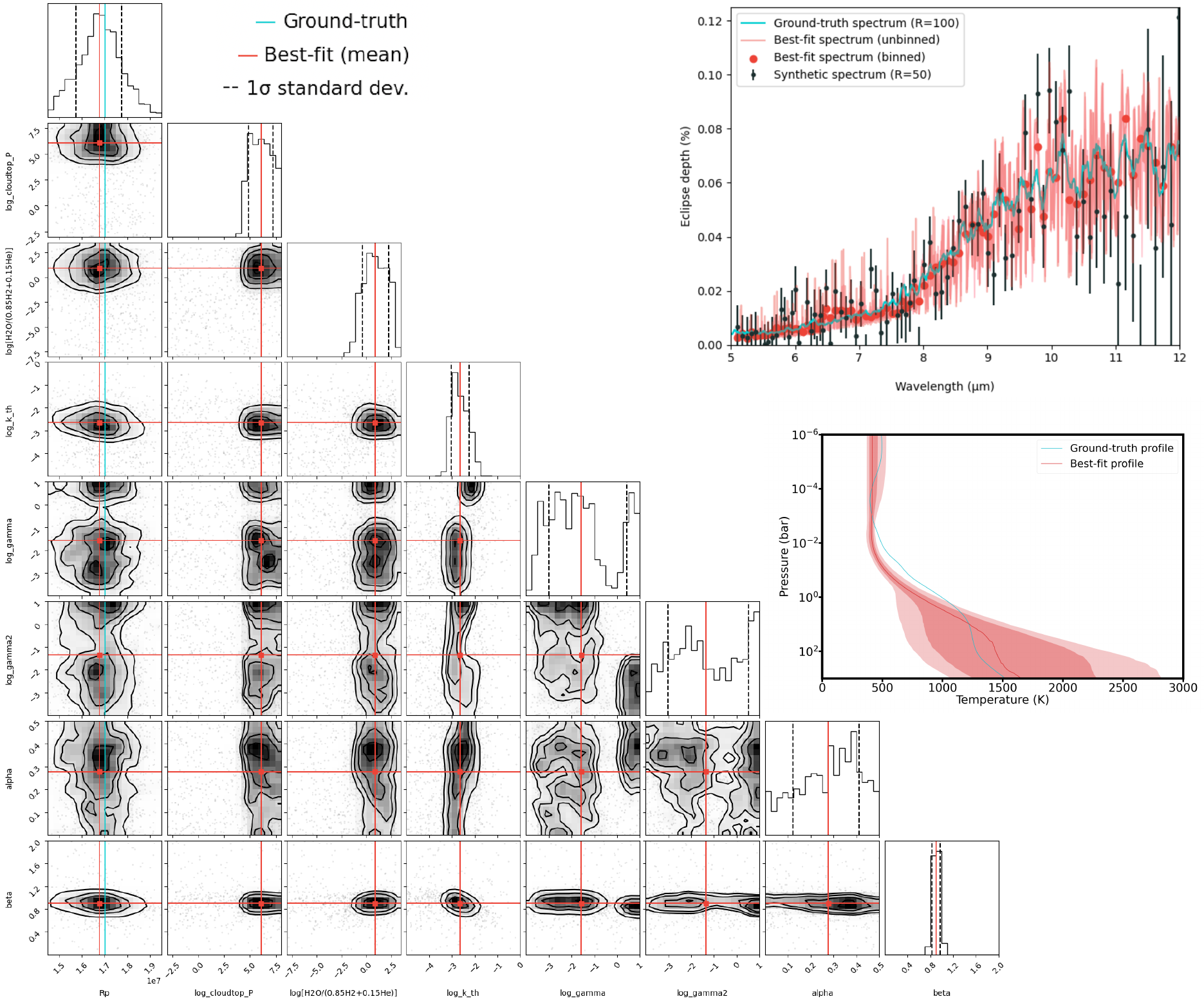}
    \caption{Same as Figure~\ref{fig:retrieval_emiss_low_H2O}, but for a water abundance of 100\%.} 
    \label{fig:retrieval_emiss_high_H2O} 
\end{figure*}

\begin{figure*}[ht]
    \centering
    \includegraphics[width=0.98\textwidth]{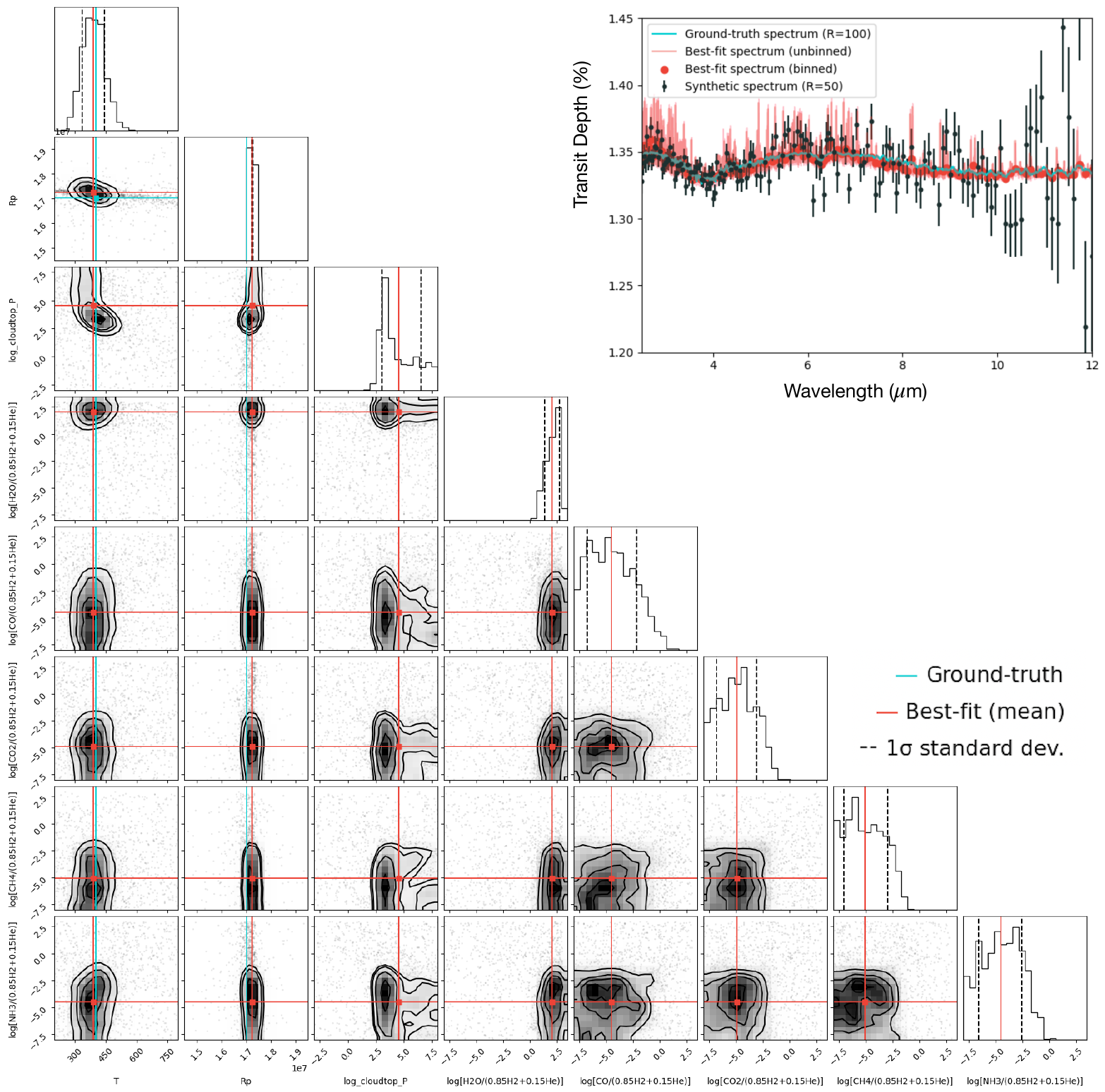}
    \caption{Same as Figure~\ref{fig:retrieval_trans_high_H2O}, but here we retrieve for the abundances of H$_2$O, CO, CO$_2$, CH$_4$, and NH$_3$ against a ``filler" background of H$_2$ and He.}
    \label{fig:retrieval_trans_all_species} 
\end{figure*}

To demonstrate the power of upcoming JWST observations to constrain the water abundances and properties of sub-Neptune exoplanet atmospheres, we run \texttt{PLATON} retrievals on several of the spectra from our model grid.  We show some representative results for transmission spectra of a GJ 1214b-like exoplanet in Figures~\ref{fig:retrieval_trans_high_H2O} and \ref{fig:retrieval_trans_low_H2O} and for emission spectra of the same planet in Figures~\ref{fig:retrieval_emiss_low_H2O} and \ref{fig:retrieval_emiss_high_H2O}.  

Transmission spectrum retrievals for a GJ 1214b-like planet have the potential to be highly constraining, due to the bright host star and large transit depth.  This can be seen in Figure~\ref{fig:retrieval_trans_high_H2O}, in which the atmospheric temperature, clear atmosphere conditions, and water abundance (retrieved here as the ratio of H$_2$O to H$_2 +$ He) are all recovered accurately.  Slight mismatches between ground-truth and retrieved parameters can be explained readily by how the retrieval model is parameterized.  For example, the upper atmosphere temperature that is sensed by the retrieval is slightly lower than the planet's equilibrium temperature (which is plotted as the ground truth value), a lower limit is obtained for the cloud-top pressure due to this being a cloud-free atmosphere, and the retrieved value of H$_2$O/H$_2$ (ground truth $= \infty$) is consistent with the upper end of our prior distribution.

Other transmission spectrum retrievals do not perform as well.  Take Figure~\ref{fig:retrieval_trans_low_H2O}, which shows a \mbox{NIRCam} retrieval for a planet with 0.1\% H$_2$O and a cloud at 0.1 mbar, in which bimodal solutions for the water abundance and cloud-top pressure are retrieved.  The ground-truth water abundance lies midway between the two bimodal peaks, but the cloud-top pressure is actually smaller than both values preferred by the retrieval.  In this case, the retrieval code attempts to fit a cloudy spectrum by either removing the water features (low H$_2$O abundance) or suppressing the scale height (high mean molecular weight), while failing to identify that the atmosphere has clouds that flatten the spectrum. This poorly-behaving retrieval comes about in part due to the specific random noise instance of the \texttt{PandExo} run, but also more importantly because of a subtle mismatch between the ``ground-truth" input model, which was calculated with \texttt{Exo-Transmit} and the \texttt{PLATON} forward model utilized by the retrieval code.  In this case, the mismatch arises due to our radiative convective equilibrium T-P profiles (\texttt{PLATON} uses isothermal profiles), differences in the water opacities used by each code, and minor discrepancies in how the cloud layer is implemented.  We note that no single one of these effects is primarily responsible for the poor quality of the retrieval.  By testing each one individually, we find it to be a subtle combination of all of these effects.  Furthermore, the retrieval is only sensitive to the subtle mismatch between the two forward models because of the very small error bars associated with the simulated GJ 1214b data 
relative to the signal size, allowing for small model differences to have significance.  

The retrieval result shown in Figure~\ref{fig:retrieval_trans_low_H2O} serves as a warning of the impact of \emph{any} slight differences between ground truth and retrieval forward model with high-precision JWST data.  In the case of any missing physics in the forward model, the retrieval code will attempt to fit the data to the best of its ability, which can sometimes result in poor recovery of the actual ground-truth atmospheric properties.  Many examples of the impact of missing physics on retrievals for JWST-quality data already exist in the literature \citep[e.g.][]{pinhas18,caldas19,changeat19,lacy20,barstow20,taylor20,cubillos21,taylor22}.  Here we highlight that a mismatch between ground-truth and retrieval arising from a combination of individually minor effects can add up to produce a major impact on the retrieved atmospheric properties.  We note that we have found similar degrees of such discrepancies across our retrievals, regardless of the resolution of the simulated data --- we ran retrievals at $R = 10$, 50, and 100 (not shown).

The emission spectrum retrievals for our GJ 1214b analog planet (Figures~\ref{fig:retrieval_emiss_low_H2O} and \ref{fig:retrieval_emiss_high_H2O}) fare much better than the transmission retrievals in accuracy but not precision, due to the smaller signal size of the secondary eclipse spectrum resulting in lower signal-to-noise.  The (relatively) larger error bars of the thermal emission spectrum are able to hide the subtle differences between the input spectrum and the retrieval code's forward model, leading to accurate recovery of the atmospheric parameters.  We recover the thermal emission degeneracies reported in Section~\ref{sec:degeneracy}.  The retrievals are unable to distinguish among high water abundances (see extended posterior distribution for the water abundance in Figure~\ref{fig:retrieval_emiss_low_H2O}), but they are readily able to rule out low water abundances in the case of a water-rich atmosphere (Figure~\ref{fig:retrieval_emiss_high_H2O}).

Finally, while our forward-modeled atmospheres do not contain any spectroscopically-active species other than H$_2$O, real atmospheres are likely to contain at least trace abundances of other molecules such as CO$_2$, CO, CH$_4$, NH$_3$, and others.  Due to the \textit{a priori} unknown composition of a typical sub-Neptune, we run an additional set of retrievals in which we attempt to simultaneously retrieve the abundances of H$_2$O, CO$_2$, CO, CH$_4$, and NH$_3$.  The results of this exercise is shown in Figure~\ref{fig:retrieval_trans_all_species} for the 100\% H$_2$O atmosphere.  Here, the high water abundance is still correctly retrieved, along with upper limits on the remaining gases, demonstrating that a water-rich atmosphere can be correctly identified from among a set of plausible absorbers.  We note however broad wavelength coverage is required for this retrieval to home in on the correct atmospheric composition.  In Figure~\ref{fig:retrieval_trans_all_species}, the simulated dataset includes wavelength coverage from both NIRSCam and MIRI.



\section{Conclusions} \label{sec:conclusion}

In this work we have presented a new grid of atmosphere models for water-rich sub-Neptune exoplanets.  The grid includes T-P profiles in radiative convective equilibrium, thermal emission spectra, and transmission spectra across a range of parameter space relevant for currently known sub-Neptunes and for calculating their thermal evolution.  The model grid is made publicly available at the following URL: \url{https://umd.box.com/v/water-worlds}.  

We include a number of novel aspects in our modeling approach for water-rich planetary atmospheres.  These include a prescription for pressure-broadening that accounts for the composition of the background atmosphere and using a non-ideal water-hydrogen-helium equation of state for calculating convective lapse rates.  

We find that these improvements to traditional exoplanet atmosphere modeling approaches primarily impact the lower atmosphere temperature structure.  This can in turn influence thermal emission spectra, especially for cases in which the planet experiences strong internal heating, which is primarily expected if the planet is very young or if it is strongly tidally heated.  Transmission spectra are not predicted to be impacted at an observable level by the updates to our modeling approach.  

We conclude that most existing widely used forward models and retrieval codes can be applied to JWST observations of sub-Neptunes without a considerable loss of accuracy, as long as the other details of the forward models (e.g.\ line lists) are accurate.  We additionally conclude that calculations of sub-Neptune internal structure and thermal evolution, which rely on atmosphere models to serve as their upper boundary conditions, are are likely to be influenced by the updated modeling treatments applied in this paper because of the considerable changes to bottom-of-atmosphere temperature conditions.  

Finally we find that thermal emission spectra are more sensitive to differences in a planet's water abundance for trace quantities of water vapor.  At high water abundance ($\gtrsim 10 \%$), the spectra become more highly degenerate against one another.  In contrast, cloud-free transmission spectra are more highly degenerate at low (near-solar) water abundance, but clouds can enhance the degree of degeneracy by erasing spectral features even for larger scale height atmospheres.  These degeneracies are seen in our mock retrievals of a GJ 1214b-like sub-Neptune exoplanet.  We furthermore uncover significant mismatches between ground-truth input values and retrieved parameters for some simulations. These discrepant results stem from subtle mismatches between the input physics between the forward model that generated the synthetic data and the retrieval code's forward model in the case of the synthetic data having very small (10's of ppm) error bars.  We caution that this situation is likely to arise with JWST data for large planets and/or bright host stars, resulting in potentially biased retrieval results that could be caused by any number of minor pieces of missing physics in the retrieval models.

Water is a key molecule that is sought out in exoplanet observations as a tracer of planet formation.  It has multiple clear and obvious spectroscopic features.  In the JWST era we expect to  measure atmospheric water abundances for a large number of exoplanets, including pushing into the sub-Neptune regime for the first time.  Models such as the ones presented in this paper are critical for interpreting atmospheric water abundance for this important population of planets.  By precisely measuring the water content of many sub-Neptune exoplanet atmospheres with JWST, we hope to ultimately address fundamental questions about how and where these ubiquitous planets typically form and evolve.  

\vspace{5mm}

\noindent We acknowledge funding from the NSF CAREER program (grant \#1931736), the Research Corporation's Cottrell Scholar program, and the NASA Habitable Worlds Program (grant \#80NSSC19K0314).  This work made use of the Deepthought2 high performance computing cluster at the University of Maryland.  We would like to thank the anonymous referee who provided useful comments and feedback on this paper.

%




\clearpage

\appendix

\section{Water opacity treatment}
\label{app:water_opac}

\begin{figure}[tb]
\plottwo{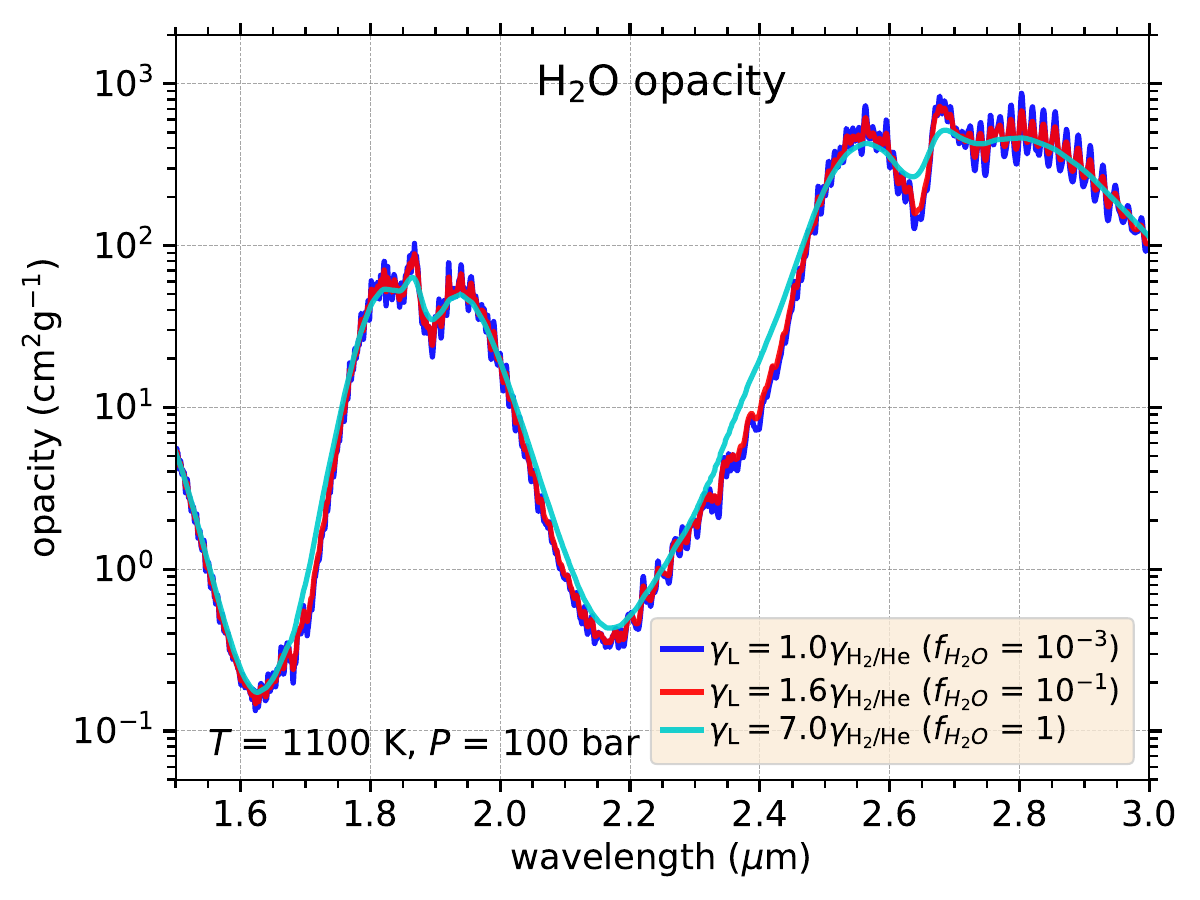}{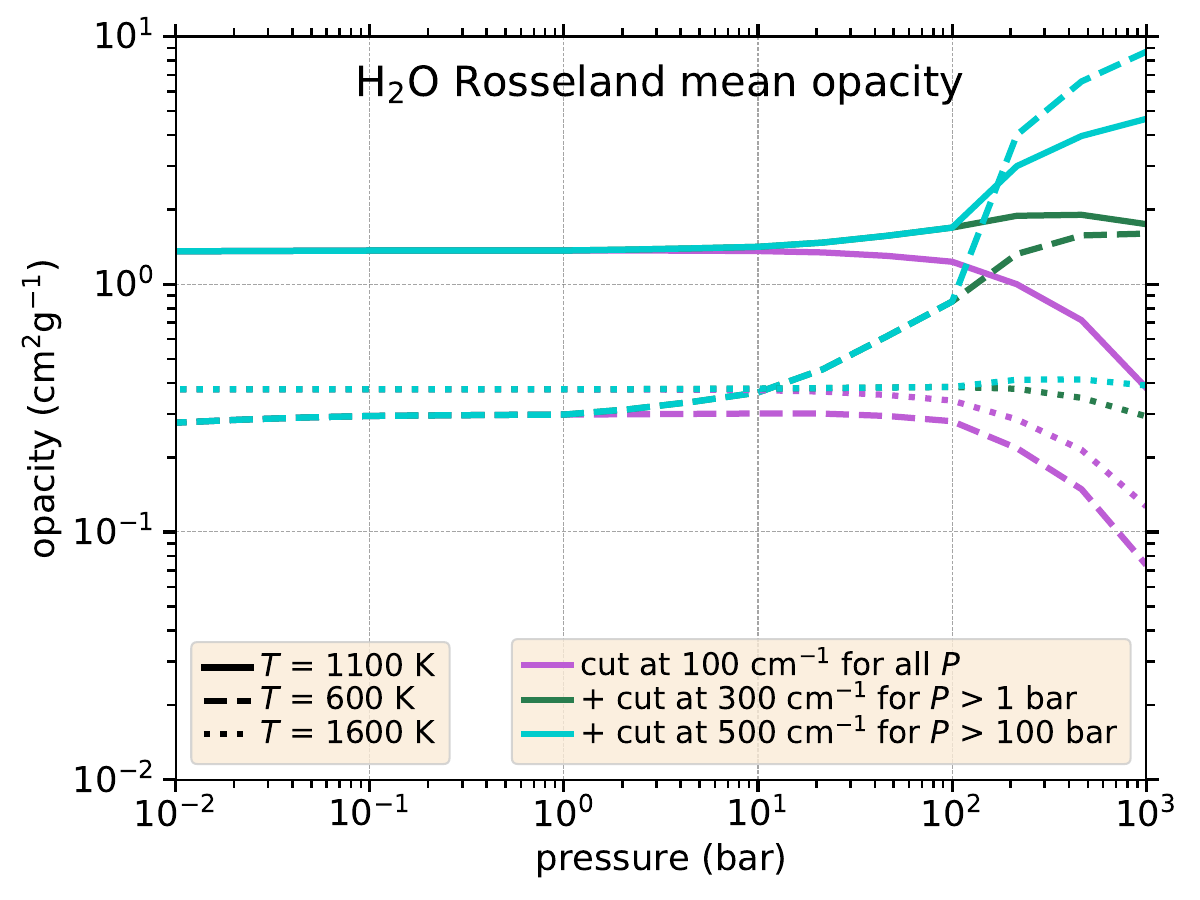}
\caption{{\bf Left:} Water opacity versus wavelength for different broadening coefficients, each case corresponding to a certain atmospheric water mixing ratio. With increasing broadening strength the spectral lines become wider and flatter resulting in a smoother opacity curve. {\bf Right:} Rosseland mean opacity versus pressure shown at three different temperatures. In contrast to the opacity calculated using a fixed line wing cut-off at 100 cm$^{-1}$ for all pressures (magenta) or calculated with extended wings up to 300 cm$^{-1}$ for $P >$ 1 bar (green), only opacity using an additional extension of the line wings to 500 cm$^{-1}$ for $P >$ 100 bar (cyan), shows the physically expected trend of constant or increasing Rosseland mean opacity with pressure throughout the modeled pressure range. Hence, the water opacity in this work is calculated using the last prescription.}
\label{fig:opac_h2o}
\end{figure}

There is no presently available water opacity database that includes both H$_2$ and self-broadening parameters.  For example, the standard broadening coefficients supplied with the POKAZATEL line list \citep{polyansky18}, which we use in this work, only consider H$_2$ and He as collision partners, and HITEMP \citep{rothman10} only includes self and air-broadening. However, the atmospheric compositions in this work encompass a broad range of water mixing ratios, from trace up to unity. For this case, \citet{gharib19} found that the Lorentzian half-width at half-maximum (HWHM) for the self-broadening of water can be approximated by multiplying the HWHM for H$_2$/He broadening, $\gamma_{\rm H_2/He}$, by a factor of 7 (see their Table 1). We calculate the Lorentzian HWHM for different water mixing ratios, $f_{\rm H_2O}$, by linearly interpolating between the H$_2$/He and pure H$_2$O limits so that
\begin{equation}
\label{eq:gamma}
\gamma_{\rm L} (f_{\rm H_2O}) = [7f_{\rm H_2O} + (1 - f_{\rm H_2O})]\gamma_{\rm H_2/He} .
\end{equation}
For very small $f_{\rm H_2O}$, $\gamma_{\rm L} \approx \gamma_{\rm H_2/He}$ and thus we set $\gamma_{\rm L} = \gamma_{\rm H_2/He}$ if $f_{\rm H_2O}  \leq 10^{-2}$ in order to keep the computational costs of the opacity calculations manageable. For the same reason, we also simplify $\gamma_{\rm L}(10^{-1.8}) = \gamma_{\rm L}(10^{-1.6})$. Fig.~\ref{fig:opac_h2o}, left panel, shows the opacity function of water for different broadening coefficients, with each case corresponding to a certain atmospheric water mixing ratio. With increasing broadening strength the spectral lines become wider and flatter resulting in a smoother opacity curve. Recall that the line strength is independent of the line profile and so the total opacity integrated over all wavelengths should be constant when varying the Lorentzian HWHM. 

As a by-product of the water opacity calculations, we provide Rosseland mean opacity tables over a wide temperature and pressure range, useful for a variety of atmospheric calculations from terrestrial planets to hot Jupiters (\url{https://umd.box.com/v/water-worlds-rosseland}). Fig. \ref{fig:opac_h2o}, right panel, shows an excerpt from the Rosseland mean opacity, displayed as function of pressure $P$. Note, since the Rosseland opacity is weighted by the inverse of the wavelength-varying opacity, two trends should be visible in the deep atmosphere where pressure broadening is important. First, the more strongly broadened cases, with diminished troughs between the spectral lines, are expected to result in higher Rosseland opacity values. Analogously, for the same reason, at a given temperature the Rosseland opacity should also increase with increasing pressure. Worryingly, calculating opacities with a line wing cut-off at 100 cm$^{-1}$ from the line center leads to a Rosseland opacity that strongly decreases with pressure for $P \gtrsim$ 10 bar. This indicates that a cut-off at 100 cm$^{-1}$ is likely too constraining, cutting off a significant region of the line wing and leading to a decrease in the overall line strength as the line becomes increasingly broader with pressure. At high temperatures, $T \gtrsim $ 1500 K, even incorporating the line cut-off prescription of \citet{gharib19}, i.e., cut-off at 300 cm$^{-1}$ for $P >$ 1 bar, results in a decreasing Rosseland mean opacity with pressure for $P \gtrsim $ 100 bar. Hence, for our final opacity tables that are used in this work, we opt for the \citet{gharib19} line wing prescription, but additionally extend the lines to 500 cm$^{-1}$ for $P >$ 100 bar. With this prescription, the Rosseland opacity is either roughly constant or steadily increases with pressure for the whole modeled pressure range, which we deem the physically expected behavior. We note that we do not include any additional water ``continuum" opacity \citep[e.g.][]{anisman22}, as this is mostly already accounted for by our inclusion of the far line wings.

\section{Calculation of the adiabatic lapse rate}
\label{app:adiabatic}

The adiabatic lapse rate is given by the adiabatic coefficient $\kappa$ (also called $\nabla_{\rm ad}$ in stellar astrophysics), defined as
\begin{equation}
\kappa \equiv \left(\frac{\partial \log T}{\partial \log P}\right)_S = \frac{P}{T} \left(\frac{\partial T}{\partial P}\right)_S,
\end{equation}
with $T$, $P$ and $S$ being the temperature, pressure and entropy per unit mass, respectively. Applying the triple-product rule $\kappa$ can be calculated from the entropy as
\begin{equation}
\kappa = -\frac{P}{T}\frac{(\partial S / \partial P)_T}{(\partial S / \partial T)_P} = - \frac{(\partial \log S / \partial \log P)_T}{(\partial \log S / \partial \log T)_P}.
\end{equation}

 The total entropy is dependent on the composition of the atmosphere and is given by the weighted sum of the constituent entropies and the mixing entropy $S_{\rm mix}$. In this work the modeled atmospheres consist of hydrogen, helium and water, so the total entropy reads
\begin{equation}
\label{eq:S}
S = \chi_{\rm \cH}S_{\rm \cH} + \chi_{\rm \cH e}S_{\rm \cH e} + \chi_{\rm \cH_2O}S_{\rm \cH_2O} + S_{\rm mix} ,
\end{equation}
where $\chi$ is the respective mass mixing ratio of the total atmospheric hydrogen ($\cH$), helium ($\cH$e) and water ($\cH_2$O) contents, each with its own set of considered species. For the hydrogen and helium quantities we refer to the equation of state (EOS) from \citet{saumon95}, hereafter SCvH95, and for water we use the IAPWS-95 EOS \citep{wagner02}. For $T > 1273$ K we extrapolate the IAPWS-95 EOS.  (IAPWS has been shown to behave well during extrapolation --- i.e., no oscillations, asymptotic behavior and tends toward the ideal gas in this high-temperature-low-pressure regime.)  
As the EOS calculations account for the mixing of the constituent species of the respective system, they naturally include the corresponding mixing entropy. For instance, the tabulated entropy of the hydrogen EOS in SCvH95, includes the species H$_2$, H, H$^+$ and e$^-$ stemming from hydrogen ionization. In practice, instead of Eq.~(\ref{eq:S}), the equation
\begin{equation}
S = \chi_{\rm \cH}[S_{\rm \cH} + S_{\rm mix, \cH}] + \chi_{\rm \cH e}[S_{\rm \cH e} + S_{\rm mix, \cH e}] + \chi_{\rm \cH_2O}[S_{\rm \cH_2O} + S_{\rm mix, \cH_2O}] + S_{\rm mix, resid}
\end{equation}
is used, with the bracket terms showing the quantities from EOS data. Note these data are pre-tabulated on a pressure and temperature grid and one needs to make a choice on how to combine them in the final model. In interior modeling a common choice is the additive volume rule so that $1/\rho(P,T)=\sum_i \chi_i/\rho_i(P,T)$ summing over all the gas species $i$, as done by e.g., SCvH95. In atmospheric sciences, the mixing of gas species is usually done via Dalton's law of additive pressures \citep{dalton1802} so that $\sum_i P_i = P$, which is also our chosen method. For instance, when calculating the entropy at 1 bar for a gas mixture with a $10^{-3}$ water fraction, we use the tabulated entropy of water at 1 mbar. 

In order to calculate the residual mixing entropy
\begin{equation}
S_{\rm mix, resid} = S_{\rm mix} - \chi_{\rm \cH}S_{\rm mix, \cH} - \chi_{\rm \cH e}S_{\rm mix, \cH e} - \chi_{\rm \cH_2O}S_{\rm mix, \cH_2O} ,
\end{equation}
one can use the general expression for the mixing entropy
\begin{equation}
\frac{S_{\rm mix}}{k_{\rm B}} = N_{\rm tot} \ln N_{\rm tot} - \sum_i N_i \ln N_i ,
\end{equation}
where $k_{\rm B}$ is the Boltzmann constant, $N_{\rm tot}$ is the total number of particles and $N_i$ the number of particles belonging to species $i$. Applying this formula to the total mixing entropy and the mixing entropies of the $\cH$, $\cH$e and $\cH_2$O subsystems leads to
\begin{equation}
\begin{split}
S_{\rm mix, resid} = - \frac{k_{\rm B}}{\overline{m}} (&f_{\rm \cH} \ln f_{\rm \cH} + f_{\rm \cH e} \ln f_{\rm \cH e} + f_{\rm \cH_2O} \ln f_{\rm \cH_2O} \\
&+ f_{\rm e} \ln f_{\rm e} - f_{\rm e}^{\cH} \ln f_{\rm e}^{\cH} - f_{\rm e}^{\cH e} \ln f_{\rm e}^{\cH e} - f_{\rm e}^{\cH_2O} \ln f_{\rm e}^{\cH_2O}) ,
\end{split}
\end{equation}
where $\overline{m}$ is the mean particle mass, $f_X$ is the volume mixing ratio of subsystem $X$ and $f_{\rm e}^X$ is the mixing ratio of electrons stemming from ionization in subsystem $X$. By construction of the atmospheric grid, the mixing ratios of hydrogen, helium and water are constant so that the only non-zero contribution to the derivative of $S_{\rm mix, resid}$, which is the quantity needed for the calculation of $\kappa$, comes from the electron terms. In this work, we neglect all additional terms associated with the dissociation and ionization of species (apart from the terms already included in the pre-tabulated data from SCvH95 and IAPWS-95) and, consequently set the electron terms to zero. Note that this approximation renders our calculation of $\kappa$ less accurate in the $P$, $T$ regime where dissociation and ionization is relevant. At lower pressures, $P \lesssim 1$ bar, dissociation becomes non-negligible (i.e., the fraction of H exceeds 1$\%$) when $T \gtrsim 2300$ K, and in the deep layers, $P \sim 100$ bar, this happens for $T \gtrsim 2800$ K. In our modeling, such temperatures are only reached for the highest internal temperature of 400 K and only at the bottom of the modeled atmosphere. With the electron terms set to zero, the derivative of $S_{\rm mix, resid}$ vanishes and is absent in subsequent equations.

The derivative of the entropy with respect to $Y \in [T, P]$ is immediately given by
\begin{equation}
\label{eq:partial_S}
\frac{\partial S}{\partial Y} = \chi_{\cH} \frac{\partial [S_{\rm \cH} + S_{\rm mix, \cH}]}{\partial Y} + \chi_{\cH e}\frac{\partial [S_{\rm \cH e} + S_{\rm mix, \cH e}]}{\partial Y} + \chi_{\cH_2O}\frac{\partial [S_{\rm \cH_2O} + S_{\rm mix, \cH_2O}]}{\partial Y} ,
\end{equation}
where we have assumed that the mass mixing ratios of the subsystems are constant with $P$ and $T$. 
This assumption follows from approximating the mean particle masses in the subsystems with $\overline{m}_{\cH} = m_{\rm H_2}$, $\overline{m}_{\cH e} = m_{\rm He}$ and $\overline{m}_{\cH_2O} = m_{\rm H_2O}$, a consequence of neglecting dissociation and ionization of species. As the volume mixing ratio $f$ for each subsystem $X$ is constant with $P$ and $T$, the mass mixing ratio, given by
\begin{equation}
\chi_X = \frac{\overline{m}_X f_X}{\sum_{i=[\cH, \cH e, \cH_2O]} \overline{m}_i f_i},
\end{equation}
is constant as well.
In logarithmic form Eq.~(\ref{eq:partial_S}) becomes
\begin{equation}
\frac{\partial \log S}{\partial \log Y} = \chi_{\cH} \frac{S_{\cH}}{S}\frac{\partial \log S_{\cH}}{\partial \log Y} + \chi_{\cH e} \frac{S_{\cH e}}{S}\frac{\partial \log S_{\cH e}}{\partial \log Y} + \chi_{\cH_2O} \frac{S_{\cH_2O}}{S}\frac{\partial \log S_{\cH_2O}}{\partial \log Y} .
\end{equation}
Note that here we omit the explicit notation of the mixing entropies of the subsystems for clarity, but these mixing entropies should be understood to be included in the respective entropies.

\begin{figure}[tb]
\plotone{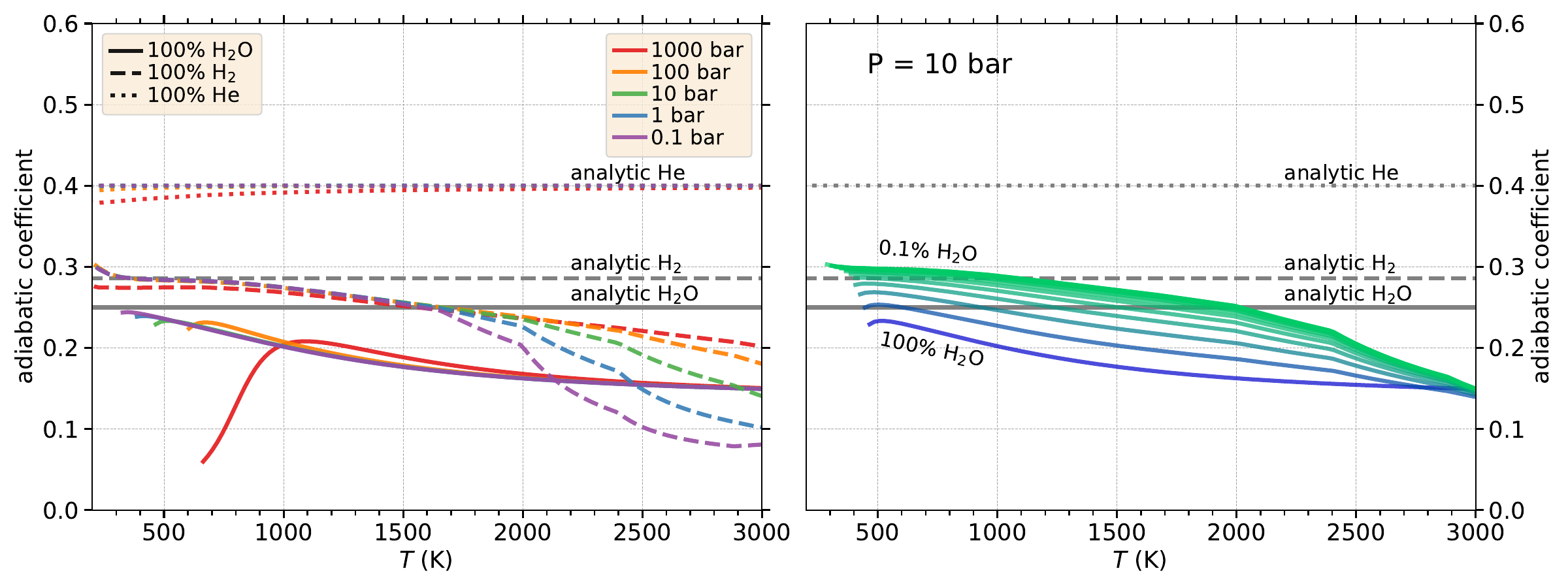}
\caption{The adiabatic coefficient versus temperature for different atmospheric compositions. {\bf Left:} The adiabatic coefficient for pure hydrogen (dashed), pure helium (dotted) and pure water (solid) atmospheres at different pressures. Gray horizontal lines show the value from the perfect-gas relation for each case. At the lower temperature end, where water is liquid or solid, the adiabatic coefficient is undefined. {\bf Right:} The adiabatic coefficient for atmospheres considered in this work, a varying water volume mixing ratio from 0.001 to 1 and a solar H/He ratio, shown at 10 bar.}
\label{fig:delad}
\end{figure}

Fig.~\ref{fig:delad} shows the temperature dependence of the adiabatic coefficient $\kappa$ for different atmospheric compositions. The left panel depicts pure hydrogen, pure helium and pure water atmosphere cases. In the lower temperature regime $\kappa$ is well-approximated by the perfect gas relation \citep[e.g., ][]{reif09}
\begin{equation}
\label{eq:kappa_analytic}
\kappa = \frac{\gamma - 1}{\gamma} = \frac{2}{2 + f} ,
\end{equation}
where $\gamma$ is the adiabatic index and $f$ is the number of degrees of freedom of the gas particles. At a low temperature, at which only translational motion and rotational energy states are relevant, mono-atomic species (like He) have $f = 3$, diatomic species (like H$_2$) have $f = 5$ and tri-atomic species (like H$_2$O) have $f = 6$ and Eq.~(\ref{eq:kappa_analytic}) returns 0.4, 0.2857, 0.25, in the respective cases. With increasing temperature vibrational energy states become occupied and the number of degrees of freedom rises, leading to a decrease in $\kappa$. In contrast, as a mono-atomic molecule has no vibrational modes, its degrees of freedom remain constant over a wider temperature regime.
At an even higher temperature, gas molecules begin to thermally dissociate and to ionize, moving further away from perfect gas behavior. The right panel of Fig.~\ref{fig:delad} shows the $\kappa$ curves for the atmospheric compositions that are used in this work, a varying water volume mixing ratio from 0.001 to 1 and a solar H/He ratio.

Lastly, the atmospheric heat capacity, needed for the convective adjustment scheme in the radiative transfer calculation, is calculated via
\begin{equation}
c_P = S \overline{m}_{\rm mol} \left(\frac{\partial \log S}{\partial \log T}\right)_P ,
\end{equation}
where $\overline{m}_{\rm mol}$ is the mean molar mass. This leads to $[c_P] = $ erg mol$^{-1}$ K$^{-1}$ which are the required units for the application in \texttt{HELIOS}.

\bibliography{Waterworlds, bib_matej}{}
\bibliographystyle{aasjournal}



\end{document}